\documentclass[a4paper,english,aps,floats,twocolumn,showpacs,nofootinbib]{revtex4}
\usepackage{pslatex}
\usepackage[T1]{fontenc}
\usepackage{graphicx}
\usepackage{epsfig}

\usepackage{calc}
\usepackage{ifthen}
\usepackage{subfigure}

\newcommand{\bea}{\begin{eqnarray}}
\newcommand{\eea}{\end{eqnarray}}

\newcommand{\lh}{\lambda_h}
\newcommand{\ls}{\lambda_s}
\newcommand{\lhs}{\lambda_{hs}}
\newcommand{\tdeps}{\tilde{\epsilon}}
\newcommand{\tdeta}{\tilde{\eta}}

\newcommand{\tdu}{\tilde{U}}
\newcommand{\tdN}{\tilde{N}}
\newcommand{\tdze}{\tilde{\zeta^2}}

\newcommand{\nc}{\newcommand}
\nc{\renc}{\renewcommand}
\nc{\eqs}[2]{\mbox{Eqs.~(\ref{#1},\,\ref{#2})}}
\nc{\eq}[1]{\mbox{Eq.~(\ref{#1})}}
\nc{\figs}[2]{\mbox{Figs.~(\ref{#1},\,\ref{#2})}}
\nc{\fig}[1]{\mbox{Fig~.(\ref{#1})}}
\nc{\be}[1]{\begin{equation} \mbox{$\label{#1}$}}
\nc{\ee}{\vspace{0.1cm}\end{equation}}

\newcommand{\bean}{\begin{eqnarray*}}
\newcommand{\eean}{\end{eqnarray*}}

\def\GeV{{\rm \ GeV}}

\def\lae{\;^{<}_{\sim} \;} \def\gae{\; ^{>}_{\sim} \;}

\def\mso{m_{s_{o}}}

\begin{document}
\title{Gauge singlet scalar as inflaton and thermal relic dark matter}
\author{Rose N. Lerner}
\email{r.lerner@lancaster.ac.uk}
\author{John McDonald}
\email{j.mcdonald@lancaster.ac.uk}
\affiliation{Cosmology and Astroparticle Physics Group, University of
Lancaster, Lancaster LA1 4YB, UK}

\begin{abstract}

We show that, by adding a gauge singlet scalar $S$ to the Standard Model which is non-minimally coupled to gravity, $S$ can act both as the inflaton and as thermal relic dark matter.
We obtain the allowed region of the ($m_s,~m_h$) parameter space which gives a spectral index in agreement with observational bounds and also produces the observed dark matter density while not violating vacuum stability or non-perturbativity constraints.
We show that, in contrast to the case of Higgs inflation, once quantum corrections are included the spectral index is significantly larger than the classical value ($n = 0.966$ for $N = 60$) for all allowed values of the Higgs mass $m_{h}$. The range of Higgs mass compatible with the constraints is $145 \GeV \lesssim m_h \lesssim 170$~GeV. The $S$ mass lies in the range  $45 \GeV \lesssim m_s \lesssim 1 $~TeV for the case of a real $S$ scalar with large quartic self-coupling $\ls$, with a smaller upper bound for smaller $\ls$. A region of the parameter space is accessible to direct searches at the LHC via $h \rightarrow SS$, while future direct dark matter searches should be able to significantly constrain the model.

\end{abstract}

\maketitle

\section{Introduction}

Although the Standard Model (SM) of particle physics reproduces experimental results well, it does not provide a mechanism for inflation, dark matter, baryogenesis or neutrino masses. There are many Beyond the Standard Model (BSM) theories which attempt to do this. Such BSM theories usually involve a new scale between the SM scale and the Planck scale and many additional particles, with the SM then viewed as the low energy remnant of some more complete theory, which is only valid up to some cut off $\Lambda < M_p$.

An alternative philosophy is to add to the SM the minimal number of new fields that are needed to address these issues. One example is the $\nu$MSM \cite{nmMSMshap}, which is the SM extended by three singlet fermions to account for neutrino masses. In this case dark matter can be explained by a keV-scale sterile neutrino while
baryogenesis occurs via leptogenesis due to sterile neutrino oscillations \cite{ol}. Therefore neutrino masses, dark matter and baryogenesis can all be explained within a very minimal extension of the SM, although this imposes
non-trivial conditions on the sterile neutrino masses and couplings \cite{shapSym}.  A scale-invariant but very weakly-coupled scalar may also be added to serve as the inflaton \cite{nmsmInf}. Other minimal extensions of the SM include the `new minimal Standard Model' \cite{Davoudiasl} and the `minimal non-minimal Standard Model' \cite{Bij}. One motivation for considering weak-scale extensions of the SM is the idea that the hierarchy problems of non-supersymmetric particle theories can be avoided if there is only one mass scale in the effective field theory below the Planck scale \cite{shaphier}.

    Recently it has been suggested that inflation might be explained purely within the framework of the SM,
with the Higgs field itself serving as the inflaton \cite{shap3}.
This has been extensively investigated in a number of papers \cite{shap3,shap2,shap1,shap4,Barvinsky:2008ia,star,wilczek,preheating,Park:2008hz}.
This is possible if the Higgs has a large
non-minimal coupling to gravity. However, in order to account for dark matter, baryogenesis and neutrino masses, it is still necessary to extend the SM. This might be achieved by combining
Higgs inflation with the $\nu$MSM, but other extensions which are consistent with entirely weak scale particle physics could also be considered. In particular, it is well-known that stable particles with weak scale masses and electroweak strength interactions (WIMPs) produce a thermal relic density of dark matter which is naturally of the correct order of magnitude.
Therefore there is a strong motivation to extend the SM by the addition of a particle with these properties.

The aim of this paper is to propose an alternative minimally-extended version of the SM which is able to explain both the mechanism for inflation and the presence of thermal relic dark matter. To this end, we add a stable gauge singlet scalar $S$ to the SM. This is the simplest extension which obeys gauge symmetry and can account for dark matter \cite{gsdm1,mcdonald1,mcdonald2,gsdm3,bento}. A discrete $Z_2$ or a global symmetry $U(1)$ must be imposed to ensure stability of the scalars; in the former case it is natural to consider real scalars, in the latter case complex scalars. We then consider whether $S$ can serve simultaneously as a thermal relic dark matter particle and as the inflaton, producing the correct density of dark matter while at the same time obeying the observational constraints on the spectral index $n$ and other inflation observables. Effectively we are replacing the Higgs scalar of Higgs inflation by the dark matter scalar $S$. As we will show, the model has the potential to relate particle physics, dark matter detection experiments and inflation observables, a connection that will be brought into focus in the near future by the LHC, the Planck satellite and future dark matter detectors.

    During the development of this paper a closely related model was proposed in \cite{clark}. This considers the same gauge singlet scalar extension of the SM to account for dark matter, but focuses on the case of Higgs inflation. As we will discuss,  there are some differences in the results for pure Higgs inflation, the model of \cite{clark} and our model, such that it may be within the reach of imminent experiments (Planck, LHC) to rule out or distinguish between Higgs inflation models and $S$-inflation.

Our paper is organized as follows. In Sec. \ref{J E} we
introduce our model. In Sec. \ref{rad cor} we review the approach to calculating radiative corrections in this class of models and derive the renormalisation group (RG) equations. In Sec. \ref{bounds} we discuss constraints coming from stability and perturbativity of the potential and slow-roll inflation observables. In Sec. \ref{dm} we discuss $S$ as dark matter, relating $\lhs$ and $m_s$. In Sec. \ref{results} we present our results and in Sec.~\ref{conc} we discuss our conclusions. Details of the derivation of the RG equations and the calculation of the dark matter density are given in the appendices.

\section{The $S$-Inflation Model}
\label{J E}

\subsection{Jordan and Einstein Frames}

The Jordan frame is the `real world' frame, where we make measurements in a standard manner. The Einstein frame is related to this by a conformal transformation which transforms the metric (and hence all other quantities) in a field dependent way.
The usefulness of transforming to the Einstein frame is that it transforms away the non-minimal coupling to gravity, leaving the Lagrangian in a familiar form, where methods for calculating physical quantities are well known.
What we think of as a conformal transformation to the Einstein frame is actually composed of two separate parts. The first is a change of conformal frame, warping the metric, and the second redefines the fields in a convenient form. Useful discussions of conformal transformations are given in \cite{conformal} and \cite{Carloni:2009gp}.

     Our procedure is to define the theory, including all radiative corrections, in the Jordan frame. We then transform to the Einstein frame in order to calculate the spectral index $n$, tensor-to-scalar ratio $r$ and running of the spectral index $\alpha$. The two frames are equivalent at low values of the fields. Since the inflation observables are calculated when perturbations re-enter the horizon i.e. at late times when the fields are small, the results calculated in the Einstein frame are the same as if we had calculated them with the non-minimally coupled scalar field in the Jordan frame.

\subsection{Non-minimally Coupled Gauge Singlet Scalar Extension of the SM}

We define the action in the Jordan frame to be
\bea
\label{Jaction}
 S_J & = & \int \sqrt{-\!g}\,d^4\! x \Big({\cal L}_{\overline{SM}} + \left(\partial _\mu H\right)^{\dagger}\left(\partial^\mu H\right) + \left(\partial _\mu S\right)^{\dagger}\left(\partial^\mu S\right) \nonumber \\
 & &  - \frac{M^2R}{2} - \xi_h H^{\dagger}H R - \xi_s S^{\dagger}SR - V(S^{\dagger}S,H^{\dagger}H)\Big)
\eea
where $V(S^{\dagger}S,H^{\dagger}H) = V^{(0)} + V^{(1)} + \cdots$. Here
\bea
\label{Jpot}
V^{(0)}(S^{\dagger}S,H^{\dagger}H)& =& \lh \left(\left(H^{\dagger}H\right) - \frac{v^2}{2}\right)^2 + \lhs S^{\dagger}S H^{\dagger}H \nonumber \\
& & + \ls \left(S^{\dagger}S\right)^2 + \mso^2S^{\dagger}S   ~\eea 
is the tree-level potential and $V^{(1)}, V^{(2)}, ...$ are the 1-loop and higher-order quantum corrections. ${\cal L}_{\overline{SM}}$ is the Standard Model Lagrangian density minus the purely Higgs doublet terms. $\mso^2$ is the constant contribution to the total $S$ mass squared,  $m_{s}^2$, which also gains a contribution from the coupling to the Higgs. For now we consider only the physical Higgs field $h$, where $H~=~\frac{1}{\sqrt{2}} \left(\begin{array}{c} 0 \\ h + v \end{array}\right) $ and $h$ is real.
We choose the direction of inflation such that $S =  \frac{s}{\sqrt{2}}$ where $s$ is real.

    Our aim is to calculate the inflation observables $n$, $r$ and $\alpha$.  This is best done using the slow-roll approximation, which cannot easily be formulated in the Jordan frame. We will therefore make a transformation of the whole action, including radiative corrections, to the Einstein frame, redefining the fields ($s \rightarrow \chi_s,~h \rightarrow \chi_h$) to ensure canonical normalisation. We then compute the slow roll parameters in the Einstein frame, using the coupling constants which we have run (in the Jordan frame) to the appropriate scale. Quantities in the Einstein frame will be denoted by a tilde (e.g. $\tilde{g}_{\mu\nu}$). From here on we set $M=M_{p}$ (reduced Planck mass), since the correction to $M$ due to the Higgs expectation value is tiny compared with $M_{P}$.

For general $h$ and $s$ the transformation to the Einstein frame is defined by
\be{2} \tilde{g}_{\mu\nu} = \Omega ^2 g_{\mu\nu} \ee with
\be{3}\label{omegaeq} \Omega ^2 = 1 + \frac{\xi_s s^2}{M_P^2} + \frac{\xi_h h^2}{M_P^2} .\ee
The fields are redefined by
\be{4} \frac{d\chi_s}{ds} = \sqrt{\frac{\Omega ^2 + 6 \xi_s ^2s^2/M_P^2}{\Omega ^4}} \ee and
\be{3a} \frac{d\chi_h}{dh} = \sqrt{\frac{\Omega ^2 + 6 \xi_h ^2h^2/M_P^2}{\Omega ^4}}, \ee

resulting in the Einstein frame action
\bea
S_E & = & \int d^4x\sqrt{-\tilde{g}}\Big( \tilde{{\cal L}}_{\overline{SM}} - \frac{M_P^2\tilde{R}}{2} + \frac{1}{2}\tilde{\partial} _\mu \chi_h \tilde{\partial}^{\mu} \chi_h + \frac{1}{2}\tilde{\partial} _\mu \chi_s \tilde{\partial}^{\mu} \chi_s \nonumber \\
 & & + A(\chi_s,\chi_h)\tilde{\partial} _\mu \chi_h \tilde{\partial} ^{\mu} \chi_s - U(\chi_s,\chi_h)\Big)
\eea
where \be{6} A(\chi_s,\chi_h) = \frac{6\xi_s\xi_h} {M_P^2 \Omega^4}\frac{ds}{d\chi_s} \frac{dh}{d\chi_h} h s  \ee
and $$U(\chi_s,\chi_h) = \frac{1}{\Omega^4}V(s,h)$$ with
\be{7} \label{pot eq} U^{(0)}(\chi_s,\chi_h) = \frac{1}{\Omega^4} \left( \frac{\lambda_h}{4}(h^2 - v^2)^2 + \frac{\lambda_s}{4}s^4 +\frac{1}{2}\mso^2s^2 + \frac{\lhs}{4} s^2 h^2 \right). \ee 

\begin{figure}[htb]
                    \centering
                    \includegraphics[width=0.3\textwidth, angle=270]{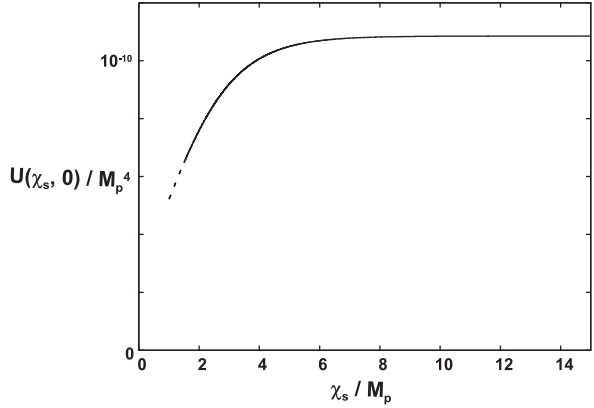}
                    \caption{\footnotesize{Classical potential in Einstein frame, in limit $s \gg M_P/\sqrt{\xi}$. This figure is plotted for real $S$ with $m_h = 160$~GeV and $\ls = 0.2$. Inflation occurs along the exponentially flat plateau.}
                    \label{treepot}}
\end{figure}

   We will be interested in inflation purely along the $s$ direction. (Inflation in the $h$ direction for real $S$ was considered in \cite{clark}.) In this case, $h = 0$,  $A(\chi_s,\chi_h) = 0$ and $\Omega ^2 = 1 + \frac{\xi_s s^2}{M_P^2}$. For $s \gg M_P/\sqrt{\xi}$, which is relevant for inflation, the classical potential in the Einstein frame becomes \cite{shap3}
\be{e6a}
\label{classpot} U^{(0)}(\chi_s,0) \approx \frac{\lambda_{s} M_{P}^{4}}{4 \xi_s^2} \left( 1 + \exp\left(-\frac{2 \chi_s}{\sqrt{6} M_{P}}
\right)\right)^{-2} ~.\ee
This is shown in Fig.~\ref{treepot}.
Thus $U(\chi_s,0) \propto 1/\xi_s^2$. Similarly, along the $h$ direction with $s = 0$, $U(0,\chi_h) \propto 1/\xi_h^2$.
Therefore if $\xi_s \gg \xi_h$, the minimum of the potential at large $s$ and $h$ will be very close
the $h = 0$ direction and so inflation will naturally occur along the $s$ direction. In the following we will consider the limit where the Higgs boson is minimally coupled to the Ricci scalar at the weak scale, $\xi_{h} = 0$, but we allow for its running by including the RG equation for $\xi_h$.

\section{Radiative corrections}
\label{rad cor}

  Our strategy is to calculate quantum corrections to the tree level potential in the Jordan frame. To do this we use the RG equations to run the couplings from the SM scale to the inflation scale. We then use those values of the coupling constants to calculate the Coleman-Weinberg correction to the potential, $V = V^{(0)} + V^{(1)}$
 \cite{Coleman:1973jx,Ford:1992mv}, where $V^{(0)}$ is given by Eq.~(\ref{Jpot}). This is then transformed to the Einstein frame to study slow-roll inflation.

\subsection{Coleman-Weinberg Potential}

   Constraints on the scalar couplings will come from the stability of the electroweak vacuum and the requirement that
the potential remains perturbative for field values less than $M_{P}$. We therefore impose the conditions for vacuum stability and perturbativity along both the $h = 0$ and $s = 0$ directions. To do this we derive the Coleman-Weinberg potential for each direction ($s$ or $h$). The $s$-direction potential is used in the slow-roll inflation calculations, while both $s$-direction and $h$-direction potentials are used to calculate the stability of the potential and to check perturbativity. We use the $\overline{\mbox{{\sc ms}}}$ renormalisation scheme throughout. The one-loop potential for the $s$ direction in the $\overline{\mbox{{\sc ms}}}$ scheme is \cite{Ford:1992mv}
$$
16\pi^2 V^{(1)}(s) = \frac{1}{4}H_s^2 \left(\ln{\frac{H_s}{\mu^2}}-\frac{3}{2}\right) +\frac{3}{4}G_s^2\left(\ln{\frac{G_s}{\mu^2}}-\frac{3}{2}\right)$$
\bea
\label{soneloop}
+ \frac{1}{4}P_s^2\left(\ln{\frac{P_s}{\mu^2}}-\frac{3}{2}\right) +  \frac{1}{4}Q_s^2\left(\ln{\frac{Q_s}{\mu^2}}-\frac{3}{2}\right)
~,\eea
where
\bea
 H_s &=& m_h^2 + \frac{1}{2}c_h\lhs s^2,~G_s = m_h^2 + \frac{1}{2}\lhs s^2, \nonumber \\
P_s &=& \mso^2 + 3c_s\ls s^2~\mbox{and}~Q_s = \left\{\begin{array}{cl} 0& \mbox{(real $S$)}\\ \mso^2 + \ls s^2& \mbox{(complex $S$).}\end{array}\right.
~.\eea
The one-loop correction for the $h$ direction is
\bea
\label{honeloop}
16\pi^2 V^{(1)}(h) &=& \frac{1}{4}H_h^2 \left(\ln{\frac{H_h}{\mu^2}}-\frac{3}{2}\right) +\frac{3}{4}G_h^2\left(\ln{\frac{G_h}{\mu^2}}-\frac{3}{2}\right)\nonumber \\
& & + \frac{1}{4}P_h^2\left(\ln{\frac{P_h}{\mu^2}}-\frac{3}{2}\right) +  \frac{1}{4}Q_h^2\left(\ln{\frac{Q_h}{\mu^2}}-\frac{3}{2}\right)\nonumber \\
& &+ \frac{3}{2}W^2\left(\ln{\frac{W}{\mu^2}}-\frac{5}{6}\right) + \frac{3}{4}Z^2\left(\ln{\frac{Z}{\mu^2}}-\frac{5}{6}\right) \nonumber \\
& & - 3T^2\left(\ln{\frac{T}{\mu^2}}-\frac{3}{2}\right)
~,\eea
where
\bea
W &=& \frac{g^2h^2}{4},~Z = \frac{\left(g^2+g'^2\right)h^2}{4},~T = \frac{y_t^2h^2}{2},\nonumber \\
H_h &=& m_h^2 + 3c_h\lh h^2,~G_h = m_h^2 + \lh h^2, \nonumber \\
P_h &=& \mso^2 + \frac{1}{2}c_s\lhs h^2~\mbox{and} \nonumber \\
Q_h &=& \left\{\begin{array}{cl} 0& \mbox{(real $S$)}\\ \mso^2 + \frac{1}{2}\lhs h^2& \mbox{(complex $S$).}\end{array}\right.
\eea
In these equations $c_s$ and $c_h$ are suppression factors to be discussed below.

\subsection{Suppression of Scalar Propagators}
\label{suppression}

       As the fields appearing in the RG equations are quantised in the Jordan frame, the commutation relation for an arbitrary scalar $\phi$
\bea
\label{comJ}
[\phi(\vec{x}), \pi(\vec{y})] = i\,\hbar\delta^3(\vec{x} - \vec{y})
\eea
is satisfied, where \cite{wilczek}
\bea
\label{piE}
\pi = \frac{\partial L}{\partial \dot{\phi}} = \sqrt{-\tilde{g}} \left(\frac{d\chi}{d\phi}\right)^2 \eta_{\mu} \tilde{g}^{\mu\nu}\tilde{\partial}_{\nu}\phi \eea
and $\eta_{\mu} = (1,0,0,0)$. This is obtained by transforming the action to the Einstein frame but not redefining the scalar fields \cite{wilczek}.  Inserting Eq. (\ref{piE}) into Eq. (\ref{comJ}) and writing in terms of the Jordan frame metric gives
\bea
[\phi(\vec{x}), \pi(\vec{y})] \equiv \Omega^2 \left(\frac{d\chi}{d\phi}\right)^2 \sqrt{-g}\;[\phi, \dot{\phi}] = i\,\hbar\delta^3(\vec{x} - \vec{y}),
\eea
and so,
$$[\phi, \dot{\phi}] = i\,\hbar c(\phi) \delta^3(\vec{x} - \vec{y})$$
where $c(\phi) = \frac{1}{\Omega^2 \left(\frac{d\chi}{d\phi}\right)^2}.$ Therefore the commutator and hence the scalar propagator will be suppressed by a factor $c(\phi)$.  In the case of minimally coupled scalars, $\left(\frac{d\chi}{d\phi}\right)^2 = \frac{1}{\Omega^2}$, so $c(\phi)~=~1$. In our case, both $s$ and $h$ are in principle suppressed by $c(s)$ and $c(h)$ respectively. In practice, we set either $c_h (\equiv c(h)) = 1$ or $c_s (\equiv c(s)) = 1$, depending on the direction of the potential being considered.  The suppression factor $c_{\phi}$ (where ${\phi}$ is $s$ or $h$) is then
\bea
\label{supeq}
c_{\phi} = \frac{1 + \frac{\xi_{\phi} {\phi}^2}{M_p^2}}{1 + (6\xi_{\phi} + 1)\frac{\xi_{\phi} {\phi}^2}{M_p^2}}.
\eea
When calculating the RG equations or Coleman-Weinberg potential, one suppression factor is inserted for each $h$ or $s$  propagator in a loop but {\em not} for the scalars corresponding to imaginary part of $S$ or the unphysical degrees of freedom of $H$. The suppression factors will have a significant effect on the running of the scalar couplings.

In the context of Higgs inflation, it was shown in \cite{burgess,barbon} that unitarity breaks down in tree-level Higgs-graviton scattering processes at energies $E \sim M_{P}/\xi_{h}$, due to the large non-minimal coupling to gravity. There are two possible ways to interpret this. In \cite{shap4} and \cite{star}, it is suggested that the apparent breakdown of unitarity should be interpreted as a change in the nature of the Higgs degree of freedom, rather than as a cut-off for new physics. In \cite{star} it was observed that at field strengths $\langle h^2\rangle \gtrsim  M_{P}^2/\xi_{h}^2$, the Higgs scalar $h$ no longer behaves as a canonically normalized scalar, resulting in suppression of the Higgs propagator as discussed above. The results for Higgs scattering cross-sections at the corresponding energies are therefore expected to be modified, which may justify the extension of the theory to energies greater than $M_{P}/\xi_{h}$.  In \cite{shap4}, it is proposed that the onset of unitarity violation could indicate a change in the dynamics of the Standard Model to a strongly coupled regime. This is described by the `chiral electroweak theory', which is equivalent to the Standard Model with the radial Higgs degree of freedom frozen\footnote{Since the chiral electroweak theory has the radial Higgs mode frozen, it explicitly breaks $SU(2)_{L} \times U(1)_{Y}$ symmetry, making it non-renormalizable \cite{shap4,clark}. We note that explicit gauge symmetry breaking is avoided when a gauge singlet scalar plays the role of the inflaton.}.

A more conservative point of view is to restrict the model to the regime where the semi-classical and adiabatic approximations are both valid \cite{burgess}. These approximations are valid when $1 \gg H/M \gg \sqrt{\ls}$, where $ H \approx \sqrt{\ls}M_P/\xi_s$ and $M \lae M_P/\xi_s$ is the scale where new physics becomes important.  As $\ls$ in our model is not constrained by phenomenology (unlike the case of Higgs inflation where $\lh$ is fixed by the Higgs mass), it would be possible to have $\lambda_s \ll 1$.

\subsection{Initial Conditions}
\label{IC}
We take the initial values of the coupling constants to be defined at the renormalisation scale $\mu = m_t$, with $m_t = 171.0$ GeV and $v = 246.22$ GeV. The gauge couplings are given by
\bea \label{ICgauge} \frac{g^2(m_t)}{4\pi} &=& 0.03344,\;\;\;  \frac{g'^2(m_t)}{4\pi} = 0.01027\;\;\;~\mbox{and}~\nonumber \\
\frac{g_3^2(m_t)}{4\pi} &= &0.1071 .
\eea
$g$ and $g'$ are obtained by an RG flow from their values at $\mu = M_Z$, which are given in \cite{pdg}, while $g_3$ is calculated
numerically. (See \cite{star} and references within for details.)

    We use the pole mass matching scheme for $\lambda_h(m_t)$ and $y_t(m_t)$ as detailed in the appendix of \cite{espinosa}. This relates the physical pole masses to the couplings in the $\overline{\mbox{{\sc ms}}}$ renormalization scheme. The remaining coupling constants are not fixed by observation and we are free to choose them. We take $\xi_h(m_t) = 0$ and choose $\xi_s(m_t)$ such that the model is correctly normalised to the COBE results at the inflation scale \cite{Lythreview,Bunn:1996py}:
$$\frac{U}{\tdeps} = (0.00271M_p)^4 .$$
$\ls(m_t)$ is not directly measurable and so we take two reasonable values: $0.2$ and $0.025$. The higher of these corresponds to $\ls(m_{t})$ close to the perturbativity limit. $\lhs(m_t)$ is treated as a free parameter although it is in principle measurable through the thermal relic $S$ dark matter density and scattering rate in dark matter detectors, as well as through the Higgs decay width to $S$ pairs should it be kinematically possible.

\subsection{Renormalisation Group Equations}

In our analysis we use the two-loop RG equations for the SM and modify these to include the leading order contributions of $S$. We also include the propagator suppression factors for the $s$ and $h$ directions.  We refer the reader to \cite{wilczek} and \cite{espinosa} for the SM one and two loop equations (including only the t-quark Yukawa coupling), reproducing here only those which are modified by the addition of the $S$ particle.
Using the technique detailed in \cite{mv} and further discussed in Appendix A,  we find for the one-loop $\beta$-functions of the scalar couplings,

\bea 16\pi^2\beta_{\lh}^{(1)} & = &  \left(18c_h^2 + 6\right)\lh^2  - 6y_t^4 \nonumber \\
& & + \frac{3}{8}\left(2g^4 + \left(g^2+g'^2\right)^2\right) \nonumber \\
 & & + \left(-9g^2 -3g'^2 + 12y_t^2\right)\lh \nonumber \\
& & +  \frac{1}{2}\left\{\begin{array}{ll} c_s^2\lhs^2 & \mbox{(real $S$)}\\ \left(1 + c_s^2\right)\lhs^2& \mbox{(complex $S$),}\end{array}\right.
\eea 

\bea 16\pi^2\beta_{\lhs}^{(1)} &= & 4c_hc_s\lhs^2 + 6\left(c_h^2+1\right)\lh\lhs \nonumber \\
& & - \frac{3}{2}\left(3g^2 + g'^2\right)\lhs + 6y_t^2\lhs \nonumber \\
& & + \left\{\begin{array}{ll}6c_s^2\ls\lhs & \mbox{(real $S$)}\\ \left(6c_s^2+2\right)\ls\lhs& \mbox{(complex $S$)}\end{array}\right.
\eea 

and \bea 16\pi^2\beta_{\ls}^{(1)} & =& \frac{1}{2}(c_h^2 + 3)\lhs^2 \nonumber \\
& & + \left\{\begin{array}{ll}18c_s^2 \ls^2 & \mbox{(real $S$)}\\ \left(18c_s^2 + 2\right)\ls^2& \mbox{(complex $S$),}\end{array}\right.
\eea 
where $\beta_{\lambda} = \frac{d\lambda}{dt}$, $t = \ln{\frac{\mu}{m_t}}$ and $\mu$ is the renormalisation scale. We choose the value of $\mu$ in order to keep the log terms in the Coleman-Weinberg potential small, setting $\mu = s_{60}$, where $s_{60}$ is the field value 60 e-foldings before the end of inflation.
In Appendix~A we relate the gauge singlet model to the matrices defined in \cite{mv} which are used to compute the RG equations.

  We also obtained the RG equations for the non-minimal couplings to one-loop. The details of this calculation are given in Appendix~A. The resulting equations are
\bea 16\pi^2 \frac{d\xi_s}{dt} &= & \left(3+c_h\right)\lhs\left(\xi_h+\frac{1}{6}\right)\nonumber \\
& & + \left(\xi_s+\frac{1}{6}\right)\left\{\begin{array}{ll}6c_s \ls & \mbox{(real $S$)}\\ \left(6c_s +2\right)\ls& \mbox{(complex $S$)}\end{array}\right.
\eea
and
\bea 16\pi^2 \frac{d\xi_h}{dt} &=& \left(\left(6+6c_h\right)\lh + 6y_t^2 - \frac{3}{2}(3g^2 + g'^2)\right)\left(\xi_h+\frac{1}{6}\right)\nonumber \\
& & + \left(\xi_s+\frac{1}{6}\right)\left\{\begin{array}{ll}c_s\lhs & \mbox{(real $S$)}\\ \left(1+c_s\right)\lhs& \mbox{(complex $S$).}\end{array}\right.
\eea

\begin{figure}[htb]
                    \centering
                    \includegraphics[width=0.3\textwidth, angle=270]{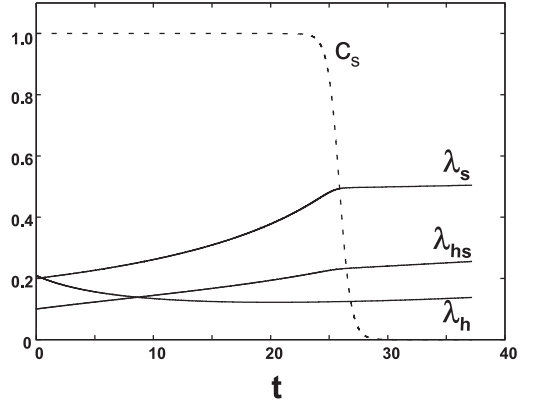}
                    \caption{\footnotesize{Running of scalar couplings showing the effect of suppressing the $s$ propagator. 
The dash line shows the corresponding value of $c_s$ when $\mu = s$. This figure is plotted for real $S$ with $m_h = 160$~GeV, $\ls(m_t) = 0.2$ and $\lhs(m_t) = 0.1$. }
                    \label{lamrun}}
\end{figure}
\begin{figure}[htb]
                    \centering
                    \includegraphics[width=0.3\textwidth, angle=270]{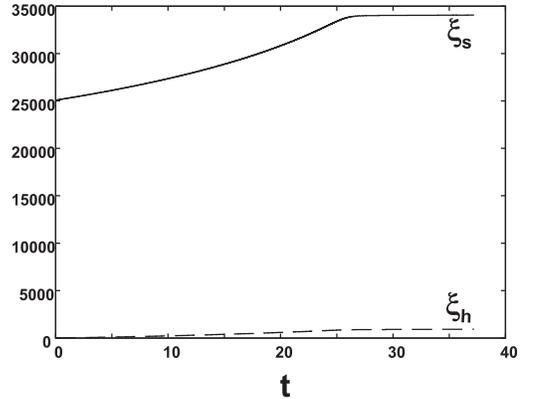}
                    \caption{\footnotesize{Running of the non-minimal couplings of $s$ and $h$ to the Ricci scalar. In this,
$\xi_h$ is set to zero at $\mu = m_{t}$.
It can be seen that $\xi_h \ll \xi_s$ throughout. The figure is plotted for real $S$ with $m_h = 160$~GeV, $\ls(m_t) = 0.2$ and $\lhs(m_t) = 0.1$.}
                    \label{nmrun}}
\end{figure}

We show in Fig.~\ref{lamrun} the running of the scalar coupling constants $\lh$, $\ls$ and $\lhs$ in the $s$ direction for the case of a real $S$ dark matter particle, assuming a small value of $\ls(m_t) = 0.025$. The figures are plotted in terms of $t = \ln{\mu/m_t}$, from $\mu = m_t$ to $\mu = M_p$. We also show the suppression factor $c_s$, Eq.~(\ref{supeq}), to demonstrate its effect on the running of the couplings. In Fig.~\ref{nmrun} we plot the running of $\xi_s$ and $\xi_h$ for the same initial conditions as in Fig.~\ref{lamrun}. We observe $\xi_h$ increasing from its initial value of 0 at $t=0$, but always remaining much smaller than $\xi_s$. This is important for the consistency of our model since inflation will occur along the $s$ direction only if $\xi_s \gg \xi_h$. Otherwise inflation would be expected to occur along a more general flat direction in the ($s,~h$) plane.

\section{Constraints}
\label{bounds}

We calculate the bounds on $m_h$ and $\lhs(m_t)$ by applying three constraints: (i) stability of the electroweak vacuum, (ii) perturbativity of the potential and (iii) consistency with the observed spectral index $n$ and with limits on the tensor-to-scalar ratio $r$ and running spectral index $\alpha$. A possible fourth constraint, `wrong-way-roll' ($\frac{dU}{d\chi_S} > 0$), which plays a role in Higgs inflation \cite{clark}, is generally not violated in our model.

\subsection{Vacuum stability and perturbativity}

We require stability of the electroweak vacuum for $s$ and $h$ up to $M_p$. (We do not consider the possibility of a metastable vacuum, which depends on the cosmological evolution of the vacuum state.) This imposes the constraints $\ls>0, \lh>0$ and either $\lhs>0$ or $\lhs^2< 4 \lh \ls$.
We will check the stability of the vacuum in both the $s$ and the $h$ direction. In practice this means that we run the RG equations with $c_s = 1$ and varying $c_h$ and again with $c_h = 1$ and varying $c_s$.

We also require the coupling constants to lie within the perturbative regime up to the Planck scale in both the $s$ direction and the $h$ direction. We apply the perturbativity condition $\lambda '_i < 4\pi$ to the coupling constants $\lambda '_i$ defined through the potential
\bea \label{primepot} V(s,h) = \frac{1}{4!}\lh'h^4 + \frac{1}{4!}\ls's^4 + \frac{1}{4}\lhs's^2h^2.
\eea
The couplings in this potential appear in the Feynman vertices without additional numerical factors.
$\lambda '_i < 4\pi$ then ensures that loop corrections are smaller than tree-level processes.
(Above $4\pi$, the coupling constants quickly grow towards a Landau pole. Therefore altering the definition of perturbativity will not significantly change our results.)
This leads to the conditions on the couplings as defined in our potential $ \lh,\;\ls < 2\pi/3$ and $\lhs < 4\pi$.

\subsection{Constraints from slow-roll inflation}

   The present observational constraints on inflation are $n = 0.960 \pm 0.013$ (1-$\sigma$), $r < 0.22$ and $ -0.068 < \alpha < 0.012$ \cite{wmap}. Inflation occurs through the standard slow-roll mechanism, which we formulate in the Einstein frame. The potential in the $\chi_s$ direction is
$$U(\chi_s) = \frac{1}{\Omega^4}\left(\frac{\ls}{4}s^4(\chi_s) + U^{(1)}(s(\chi_s))\right)  ~,$$
where $U^{(1)}(s)$ is given by Eq.~(\ref{soneloop}). The slow roll parameters are\footnote{We use $\tilde{\zeta}$ rather than
$\tilde{\xi}$ to avoid confusion.}
\bea
\tdeps &=& \frac{M_p^2}{2}\left( \frac{1}{\tdu}\frac{d\tdu}{d\chi_S}\right)^2, \nonumber \\
\tdeta &=& \frac{M_p^2}{\tdu}\frac{d^2\tdu}{d\chi_S^2} \nonumber
\eea
and
\bea
\tdze &=& \frac{M_p^4}{\tdu^2}\frac{d\tdu}{d\chi_s}\frac{d^3\tdu}{d\chi_s^3}.
\eea
From these we can calculate the observable quantities
\bea
n &=& 1 - 6\tdeps + 2\tdeta ~,\nonumber \\
r &=& 16\tdeps \nonumber
\eea
and
\bea
\alpha &=& \frac{dn}{d\ln{k}} = -16\tdeta\tdeps + 24\tdeps^2 + 2 \tdze.
\eea
The number of e-foldings of inflation is given by the standard expression \cite{Lythreview}
\be{sr1} \tdN = \int^{\chi_{\tdN}}_{\chi_{end}}\frac{1}{M_p^2}\frac{\tdu}{\frac{d\tdu}{d\chi_S}}d\chi_S  ~,\ee
where the end of inflation is defined by $\tilde{\eta} = 1$. Although this is calculated in the Einstein frame, it is straightforward to show that $\tilde{N}$ is equal to the number of e-foldings in the Jordan frame $N$ up to a small correction\footnote{$\tilde{N}$ is defined as $\ln(\tilde{a}_{end}/\tilde{a}_{N})$, where $\tilde{a}$ is the scale factor of the Friedmann-Robertson-Walker metric in the Einstein frame, which is related to the conventional scale factor by $\tilde{a} = \Omega a$.}, $ \tilde{N} \approx N + \ln(1/\sqrt{N})$. We will use $\tilde{N} = 60$ when calculating inflation observables. This is a reasonable assumption given that the reheating temperature in this model will be high\footnote{We note that $\tilde{N}=55$ gives $n=0.963$, therefore $n$ is not particularly sensitive to the value of $\tilde{N}$. We plan to compute the reheating temperature precisely in a future paper, which will fix $\tilde{N}$.}.

    Using the tree-level potential and the approximation $\frac{\xi_s s^2}{M_p^2}~\gg~1$ we estimate the tree-level
slow-roll parameters to be $\tdeps \simeq \frac{4}{3}\frac{M_p^4}{\xi_s^2 s^4}$, $\tdeta \simeq -\frac{4}{3}\frac{M_p^2}{\xi_s s^2}$ and $\tdze \simeq \frac{16}{9}\frac{M_p^4}{\xi_s^2 s^4}$, where $s_{\tilde{N}}^2 \approx 4 M_P^2 \tilde{N}/3 \xi_s$. A calculation of the tree-level spectral index, tensor-to-scalar ratio and running spectral index gives
$$ n^{(0)} \approx 1 -\frac{2}{\tdN} - \frac{3}{2 \tdN^2} + O\left(\frac{1}{\tdN^3}\right) = 0.966 ~;$$
$$ r = 3.3\times 10^{-3}  \; ; \;\;\; \alpha = 6.2 \times 10^{-4}            ~.$$
Thus $r$ and $\alpha$ are negligibly small when compared with the observational limits.

     Radiative corrections have a significant effect on the slow-roll parameters. This is not surprising, as the tree level potential is exponentially flat and the radiative corrections add a small but significant slope.
Including radiative corrections we find
\bea
\label{epsilon eq}
\tdeps  = \frac{M_p^2}{2} \left( \frac{ds}{d\chi_s} \right)^2 \left(\frac{4}{s \Omega^2} + \frac{F X}{s}\right)^2
\eea
and
\bea
\label{eta eq}
\tdeta \simeq \frac{1}{\Omega^4} \left( \frac{ds}{d\chi_s} \right)^2  \left( 48\xi_s^2 - \frac{48\xi_s^3s^2}{M_p^2} + \frac{36 \xi_s^3s^2}{M_p^2} F X    \right)
~,\eea
where
\bea
X &=& (1 + F \ln{s} + D)^{-1}, \\
F &=&\frac{1}{8\pi^2 \ls} \Big(\frac{1}{4}\left(c_h^2 + 3\right)\lhs^2 \nonumber \\
& & + \left.\left\{\begin{array}{cl} 9c_s^2\ls^2& \mbox{(real $S$)}\\ \left(9c_s^2 + 1\right)\ls^2& \mbox{(complex $S$)}\end{array}\right. \right)
\eea
and
\bea
D &=& -\frac{3}{2} F + \frac{1}{16 \pi^2 \ls}\left(\frac{c_h^2\lhs^2}{4}  \ln{\frac{c_h \lhs}{2\mu^2}} + \frac{3\lhs^2}{4}\ln{\frac{\lhs}{2\mu^2}} + \right. \nonumber \\
& & \left. \left\{\begin{array}{cl} 9c_s^2\ls^2\ln{\frac{3c_s\ls}{\mu^2}}& \mbox{(real $S$)}\\ 9c_s^2\ls^2\ln{\frac{3c_s\ls}{\mu^2}} + \ls^2\ln{\frac{\ls}{\mu^2}} & \mbox{(complex $S$).}\end{array}\right.  \right) .
\eea
The terms originating from $U^{(1)}$ are subdominant in $\tdeta$, but for a range of values of $\lhs$ and $\ls$ they can become more important than the tree-level result in $\tdeps$.

We calculate the field value at 60 e-foldings before the end of inflation as follows. First we calculate $s_{end}$ using $|\tdeta| = 1$. This gives (tree level)
\bea
s_{end}^2 \simeq \frac{4}{3} \frac{M_p^2}{\xi_s}.
\eea
Then the standard expression \eq{sr1} is integrated using Eq. (\ref{epsilon eq}) and the approximation $X$~=~constant to give
\bea
\label{s60 eq}
\tdN = \gamma \ln \left({\frac{4 + F X \Omega^2(s_N)}{4 + F X \Omega^2(s_{end})}}\right) - \frac{3}{4} \ln\left({\frac{\Omega^2(s_N)}{\Omega^2(s_{end})}}\right)
~,\eea
with $\Omega^2$ as defined in Eq.~(\ref{omegaeq}) and
\bea
\gamma = \frac{1}{2F X \xi_s} + \frac{6}{2F X} + \frac{3}{4}.
\eea

\section{Thermal relic dark matter}
\label{dm}

\begin{figure}[htb]
                    \centering
                    \includegraphics[width=0.3\textwidth, angle=270]{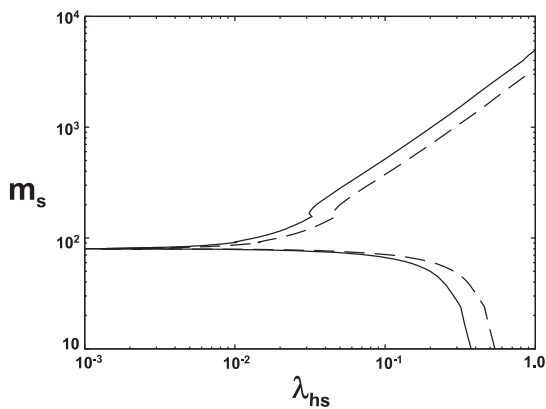}
                    \caption{\footnotesize{The value of $m_s$ as a function of $\lhs(m_{t})$ necessary to produce the correct density of thermal relic dark matter. In this example $m_h = 160.0$~GeV. The solid line indicates real $S$ and the dashed line complex $S$ scalars.}
                    \label{DMfig}}
\end{figure}

      We assume that dark matter is due to thermal relic gauge singlet scalars. The non-minimal coupling to gravity will not affect the $S$ dark matter density as the field is at very low values compared to $M_p$. If we assume that a gauge singlet scalar is responsible for the observed dark matter density ($\Omega_{DM} h^2= \rho_{DM}/\rho_c = 0.1131 \pm 0.0034$ \cite{wmap}) then we obtain a relationship between $m_s$ and $\lhs(m_{t})$. We use the Lee-Weinberg approximation \cite{leeweinberg} to calculate the relic density of $S$. This is discussed in Appendix~B, where we also review the annihilation cross-sections. For a given $\lhs$ and $m_h$ there are up to four corresponding values of $m_s$. An example is shown in Fig.~\ref{DMfig} for the case where $m_h = 160$ GeV. The cusp-like feature is due to $S$ annihilations to $WW$ and $ZZ$ pairs close to the Higgs pole. In this region the $S$ mass is relatively insensitive to $\lhs$. Note also that large values of $\lhs$ are possible for $m_{s}$ slightly below the Higgs pole.

\section{Results}
\label{results}

\begin{figure*}[htb]
  \centering
    \subfigure[\footnotesize{Real $S$, $\ls(m_t) = 0.025$}]{\label{boundsA} \includegraphics[width=0.3\textwidth, angle=270]{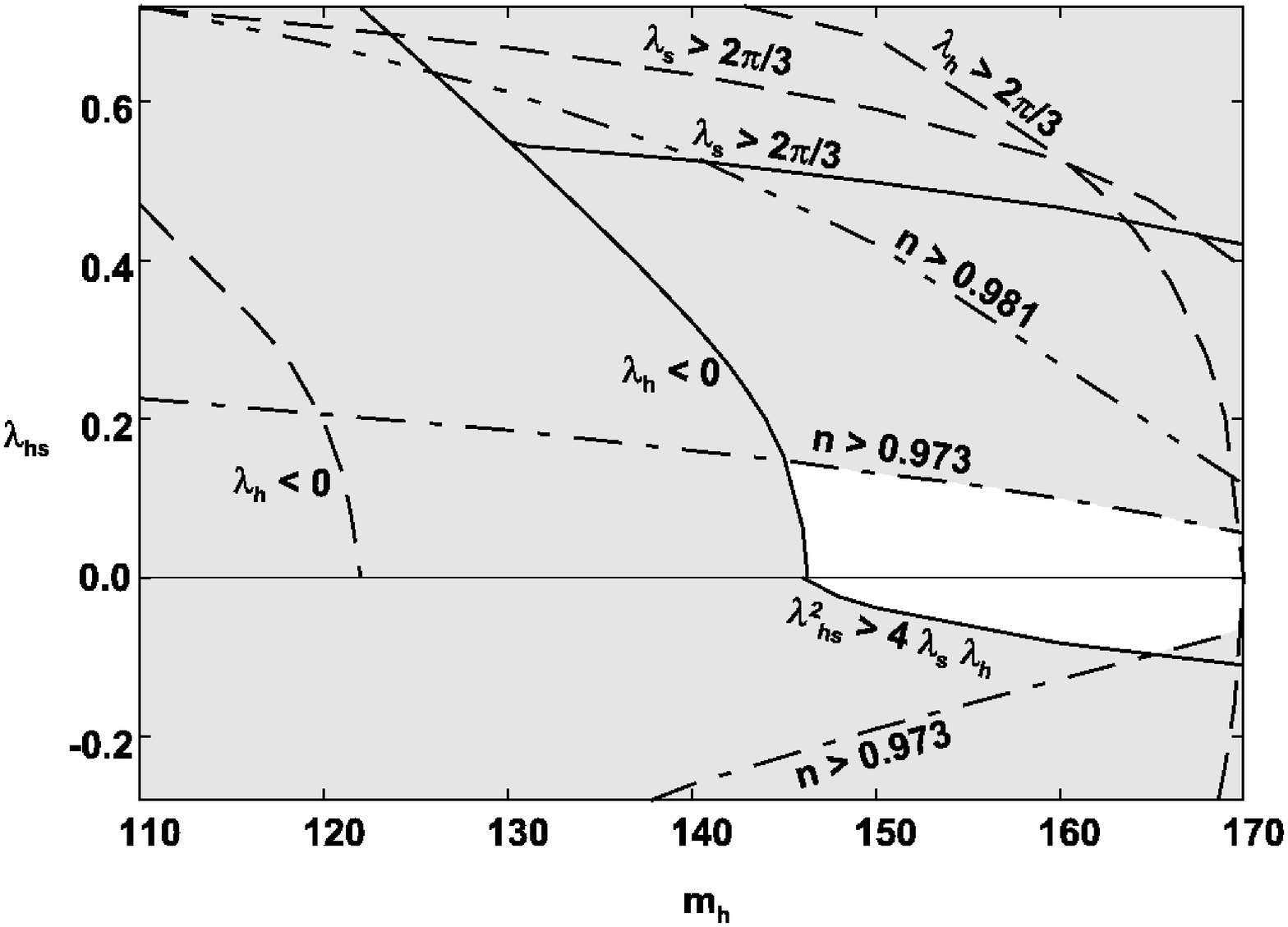}}
    \subfigure[\footnotesize{Complex $S$, $\ls(m_t) = 0.025$}] {\label{boundsB} \includegraphics[width=0.3\textwidth, angle=270]{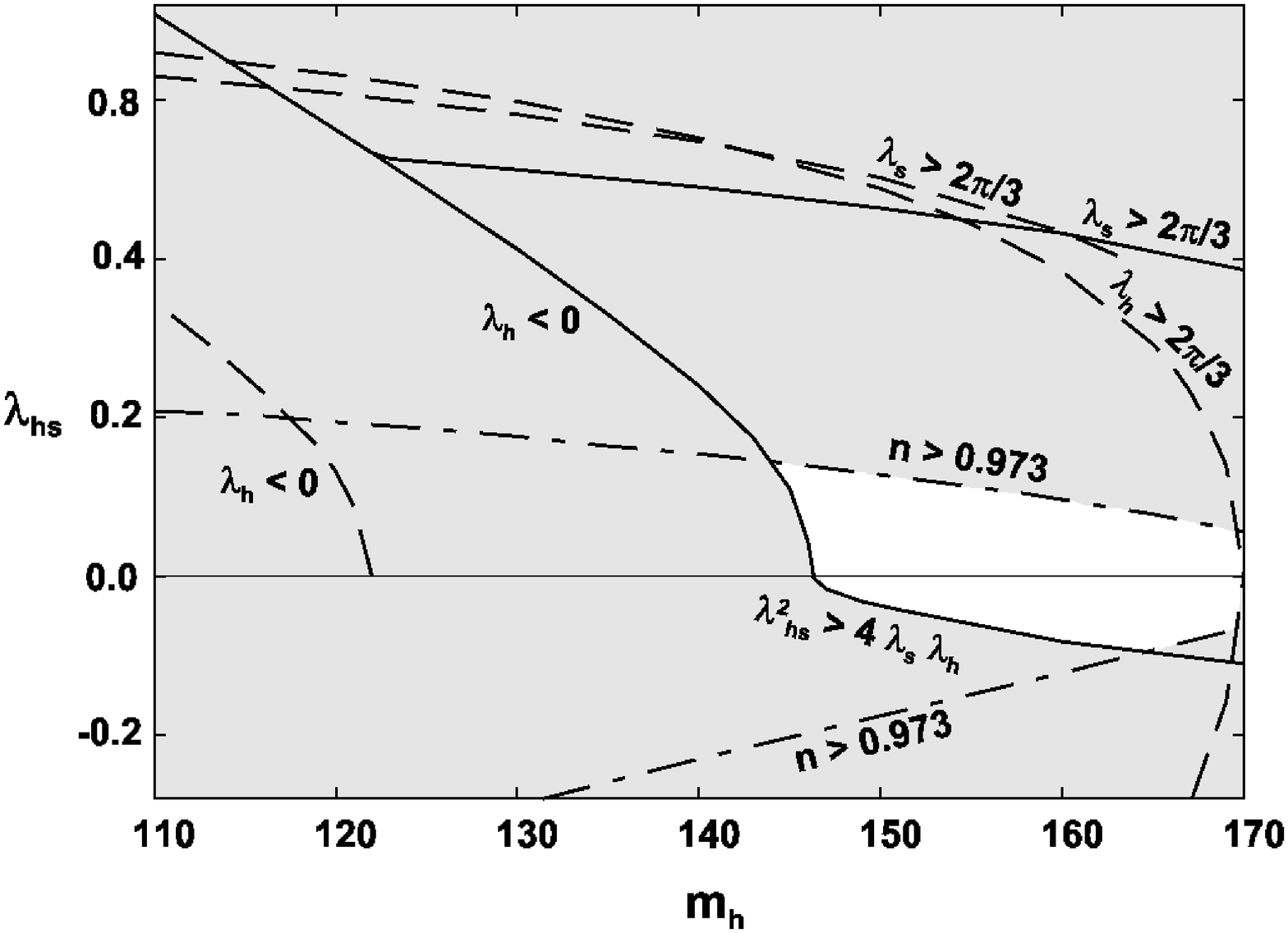}} \\
    \subfigure[\footnotesize{Real $S$, $\ls(m_t) = 0.2$}] {\label{boundsC} \includegraphics[width=0.3\textwidth, angle=270]{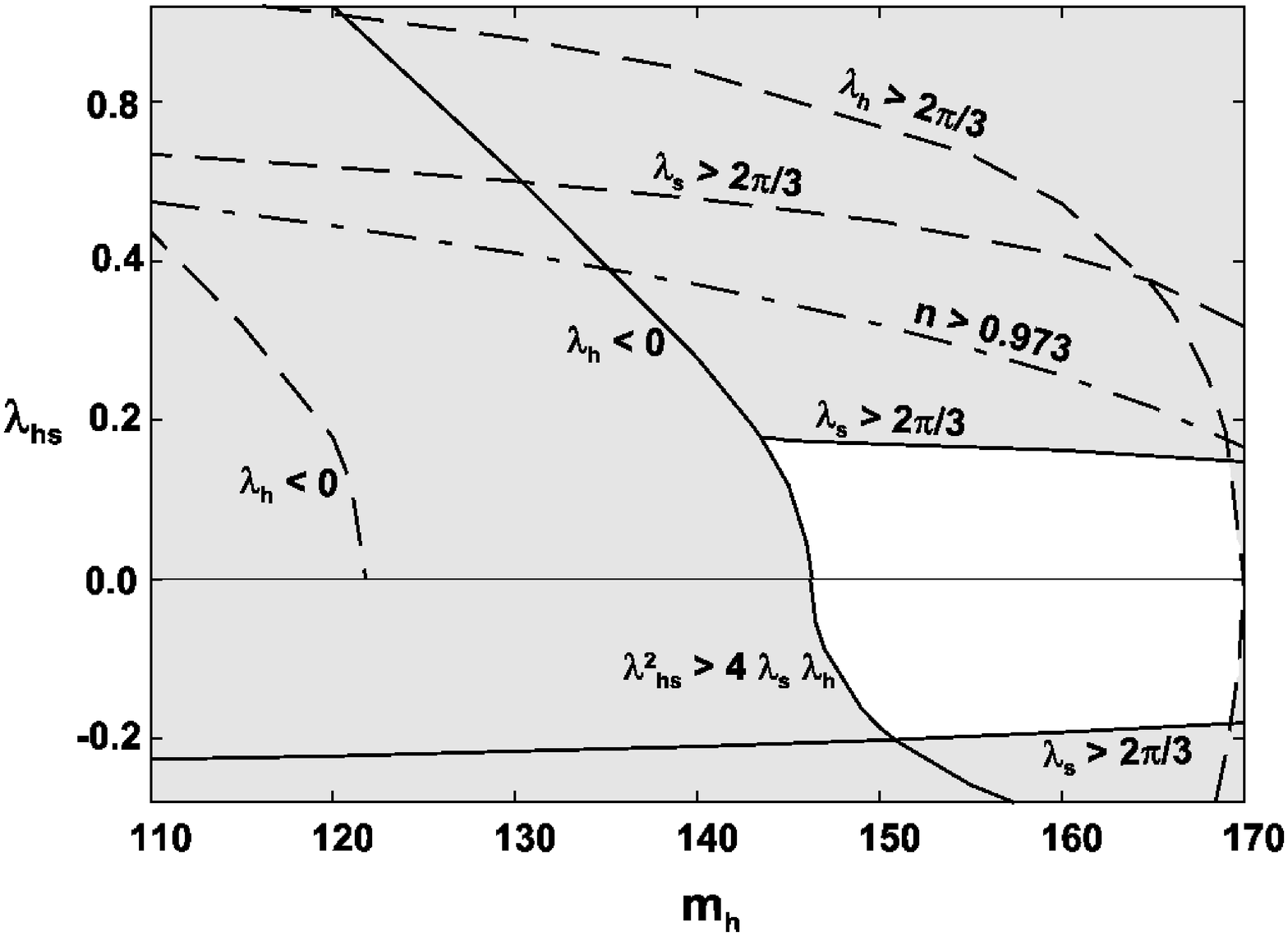}}
    \subfigure[\footnotesize{Complex $S$, $\ls(m_t) = 0.2$}] {\label{boundsD} \includegraphics[width=0.3\textwidth, angle=270]{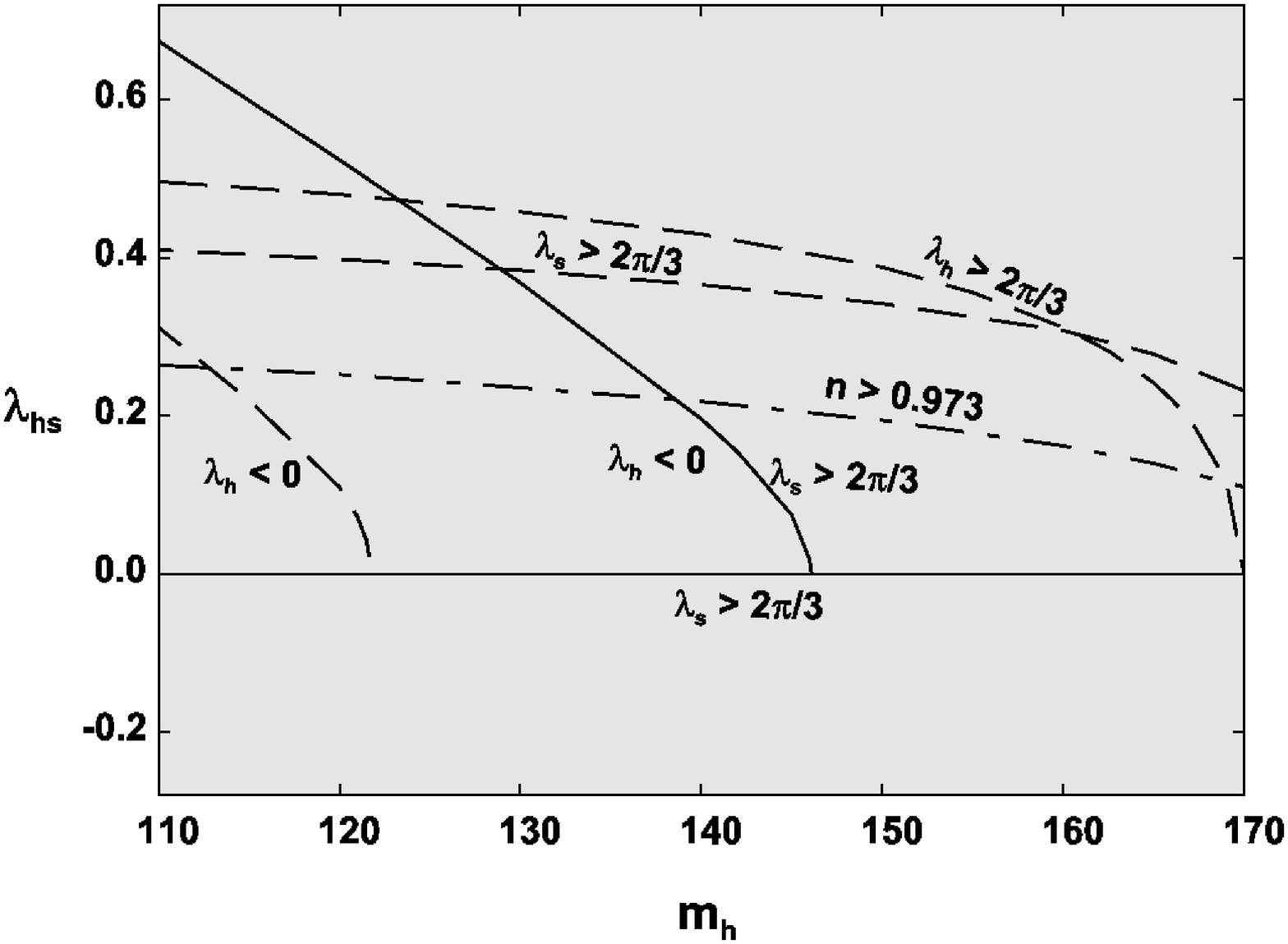}}
  \caption{\footnotesize{Allowed region for inflation in the $s$-direction. Excluded regions are shown in grey. Limits from couplings in the $s$-direction are shown with dashed lines, those from the couplings running in the $h$-direction have solid lines and the 1-$\sigma$ upper limit on $n$ is dot-dashed. In (a) we show the line $n = 0.981$ (dot-dot-dash) demonstrating the variation of $n$.}}
\end{figure*}

\begin{figure}[htb]
                    \centering
                    \includegraphics[width=0.3\textwidth, angle=270]{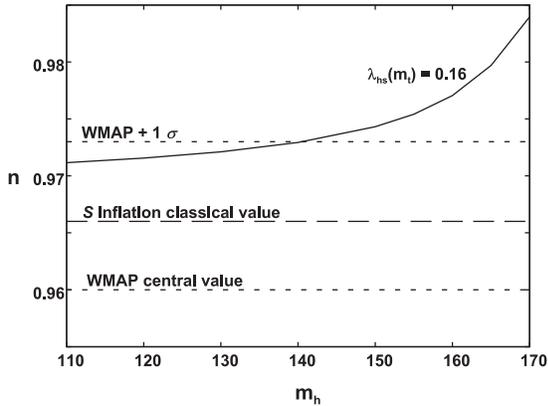}
                    \caption{\footnotesize{Showing the variation of $n$ with $m_h$ for $\ls(m_t) = 0.025$ and $\lhs(m_t) = 0.16$. WMAP central value and 1 $\sigma$ upper bound shown with short dashed lines; classical $n$ for $S$ inflation shown with dashed line.}
                    \label{varynfig}}
\end{figure}

\begin{figure*}[htb]
  \centering
    \subfigure[\footnotesize{Real $S$, $\ls(m_t) = 0.025$}]{\label{msA} \includegraphics[width=0.3\textwidth, angle=270]{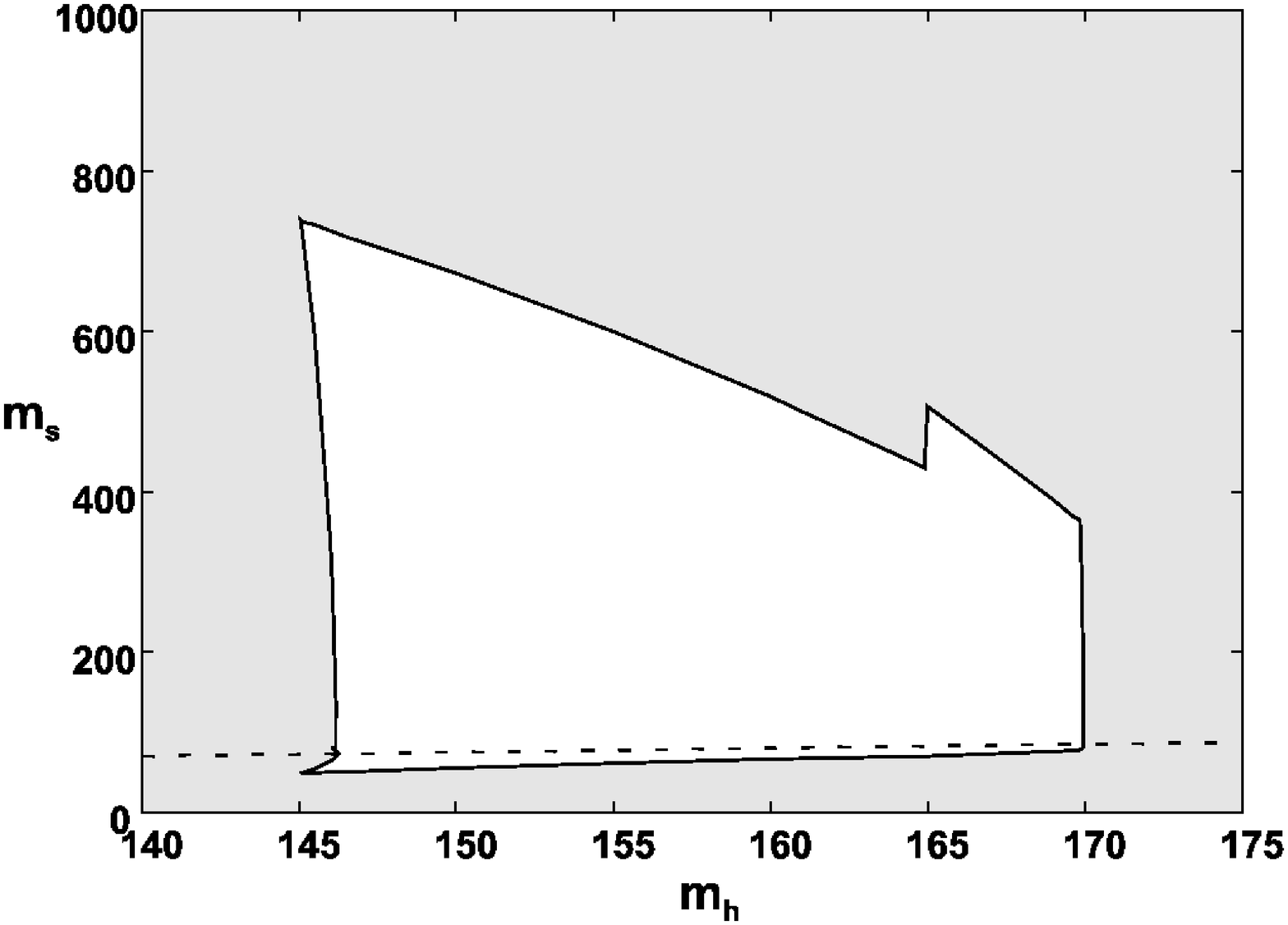}}
    \subfigure[\footnotesize{Complex $S$, $\ls(m_t) = 0.025$}] {\label{msB} \includegraphics[width=0.3\textwidth, angle=270]{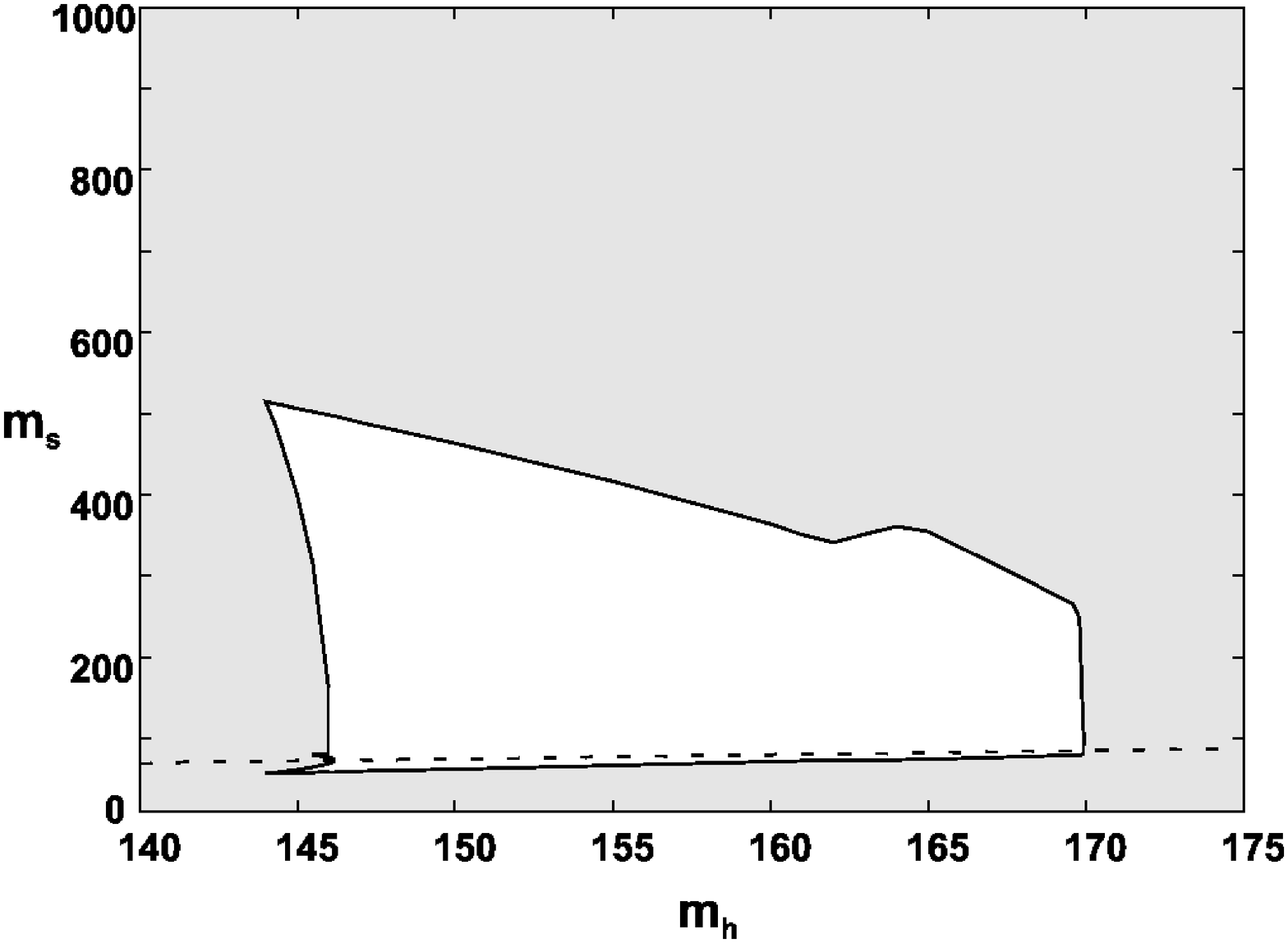}} \\
    \subfigure[\footnotesize{Real $S$, $\ls(m_t) = 0.2$}] {\label{msC} \includegraphics[width=0.3\textwidth, angle=270]{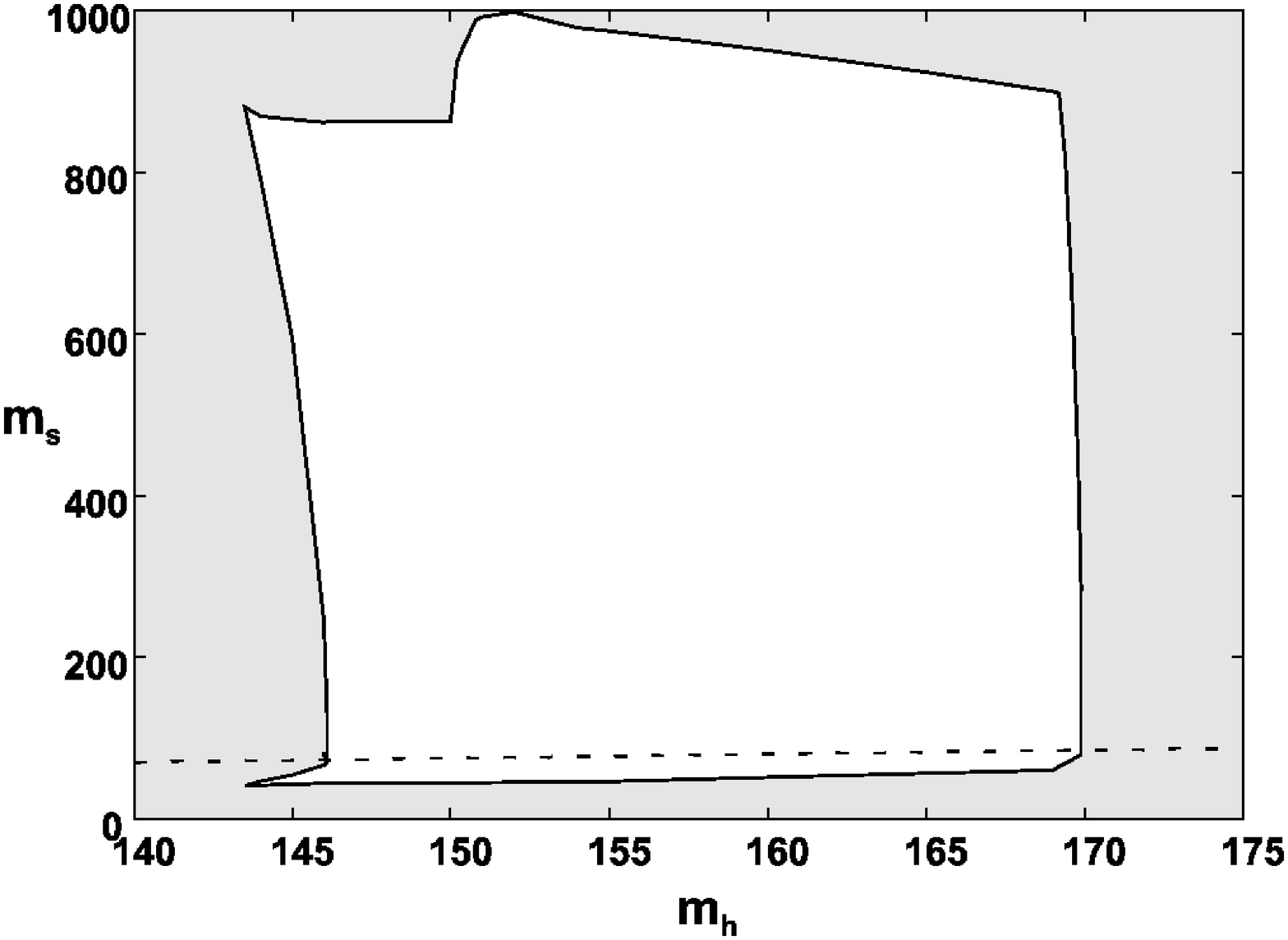}}
    \subfigure[\footnotesize{Complex $S$, $\ls(m_t) = 0.2$}] {\label{msD} \includegraphics[width=0.3\textwidth, angle=270]{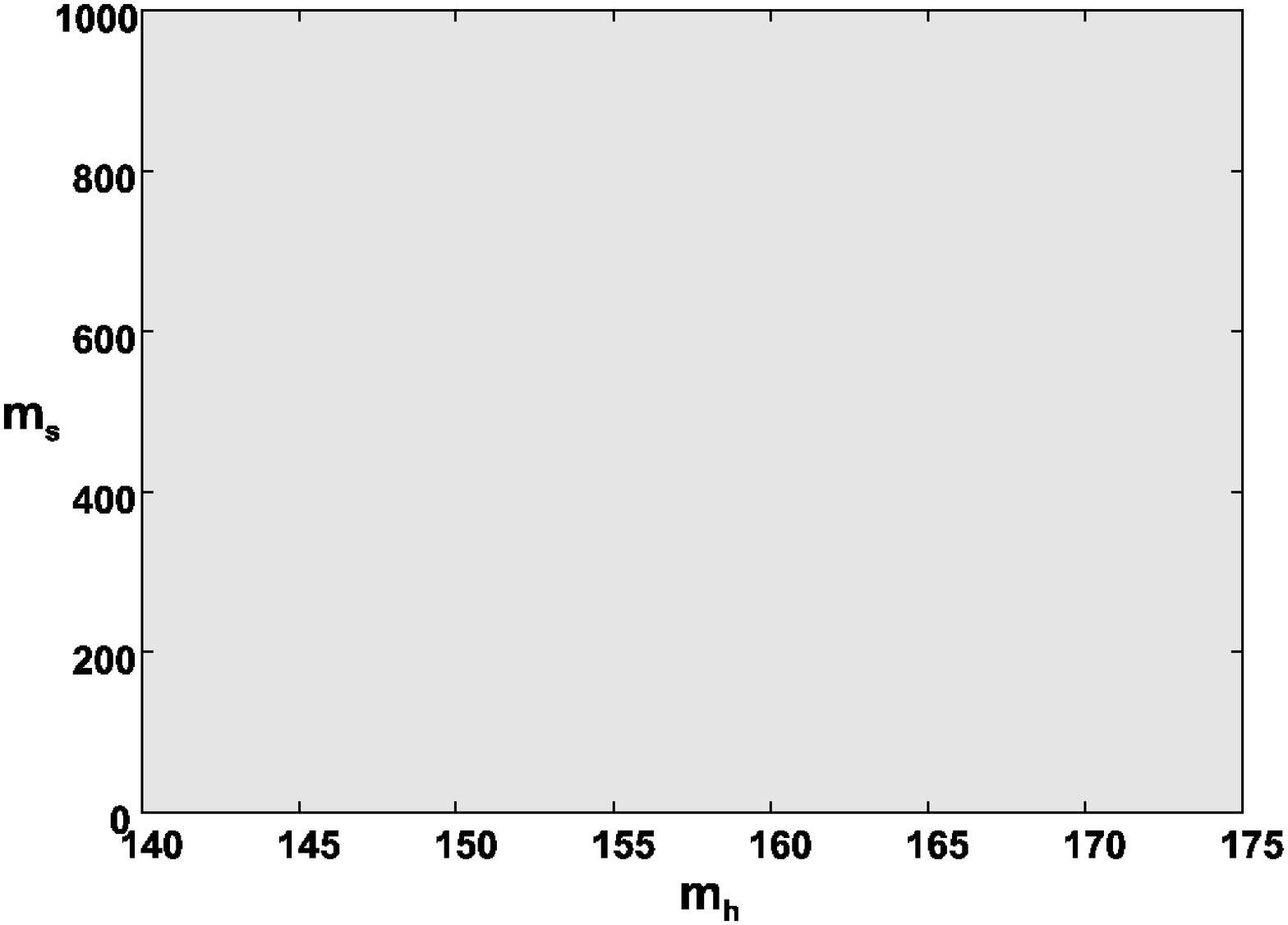}}
  \caption{\footnotesize{Allowed region for inflation in the $s$-direction, with a 1-$\sigma$ upper limit on $n$. Excluded regions are shown in grey and all masses are in GeV. The dashed line shows $m_h = 2m_s$. Below this line, production of $S$-particles at the LHC (via $h\rightarrow SS^{\dagger}$ decay) is possible. There is no allowed region in (d).}}
\end{figure*}

In Fig.~\ref{boundsA} we show the case of real $S$ with `small' $\lambda_{s}(m_{t}) = 0.025$. The range of allowed Higgs mass is $145 \GeV \lesssim m_{h} \lesssim 170 \GeV$, where the lower bound is from vacuum stability in the $h$ direction combined with 5-year WMAP 1-$\sigma$ upper bound $n < 0.973$ and the upper bound is from perturbativity of $\lambda_{h}$ in the $s$ direction.
The corresponding range of $\lhs(m_{t})$ is $|\lhs(m_{t})| \lesssim 0.15$. Larger values of $n$ allow larger $|\lhs(m_{t})|$, up to an upper bound $|\lhs(m_{t})| \approx 0.55$ (at $n \gtrsim 0.980$), which comes from the perturbativity bound on $\lambda_{s}$ in the $h$ direction. In this case the lower bound on the allowed Higgs masses is shifted downwards to $130 \GeV \lesssim m_{h} \lesssim 170 \GeV$. In Fig.~\ref{boundsB} we show the corresponding results for complex $S$. The allowed parameter space is very similar to the case of real $S$.

In Fig.~\ref{boundsC} we show the results for the case of `large' $\lambda_{s}(m_{t}) = 0.2$. In this case the range of Higgs mass is similar to the small $\lambda_{s}(m_{t})$ case, but now the origin of the bound is perturbativity of $\lambda_{s}$ in the $h$ direction rather than the WMAP upper bound on $n$. As $\lambda_{s}(m_{t})$ increases from 0.2, the allowed parameter space will rapidly diminish due to the decrease of the $\lambda_{s}$ perturbativity upper bound on $\lhs(m_{t})$. As seen in Fig.~\ref{boundsD}, the allowed parameter space vanishes for the corresponding case with complex $S$.

As discussed in Sec.~\ref{suppression}, when $\ls \ll 1$ we can avoid potential problems due to unitarity violation. Choosing a small value of $\ls$ at the scale of inflation to satisfy this bound should not be a problem. Comparing Fig.~\ref{boundsA} and Fig.~\ref{boundsC} we see the bound on $\lhs$ due to $n$ decreases with $\ls$, therefore we would expect the allowed range to decrease further with even smaller $\ls$. A smaller range of $\lhs$ will tend to drive $m_s$ closer to the Higgs pole (Fig.~\ref{DMfig}), increasing the chances that it could be detected in the near future.

An important point is that the value of $n$ can be significantly larger than the classical value $n = 0.966$ over the whole range of allowed Higgs mass.  In Fig.~\ref{varynfig} we show an example of the variation of $n$ with $m_h$ for fixed $\lhs$ and $\ls$.  This contrasts with the case of Higgs inflation without additional scalars, where a significant increase of $n$ relative to the classical value is possible only for a small range of Higgs mass close to the vacuum stability lower bound; in \cite{wilczek} a significant increase of $n$ from the classical value is obtained only for $m_{h} \lesssim 132 \GeV$. Therefore if Planck, which will measure $n$ to a 2-$\sigma$ accuracy of $\pm 0.005$, should find $n$ significantly larger than 0.966 + 0.005 while LHC finds a Higgs with mass larger than $135 \GeV$, then $S$ inflation will be compatible with the observations but Higgs inflation will be ruled out. We may also compare the range of Higgs mass allowed by $S$ inflation with that allowed by vacuum stability and perturbativity in the Standard Model. In \cite{ellisH} the range is given as $128.6 \GeV \lesssim m_{h} \lesssim 175 \GeV$, where the lower bound is from vacuum stability and the upper bound is from perturbativity of the Higgs self-coupling up to $M_{P}$. We see that the allowed range in $S$ inflation is somewhat narrower. Therefore $S$ inflation may be ruled out relative to the conventional SM if $m_{h}$ is observed close to the SM lower or upper bound.

Fig.~7 shows the range of $m_s$ and $m_{h}$ which is consistent with $S$ inflation and thermal relic $S$ dark matter when $n \leq 0.973$ and all vacuum stability and perturbativity constraints are satisfied. We also show the line $m_s = m_{h}/2$, which is the limit at which it is possible to pair produce $S$ scalars via Higgs decay at the LHC \cite{bargerR}.
For the case of real $S$ and `small' $\ls(m_{t}) = 0.025$, Fig.~\ref{msA}, we see that $m_s$ is mostly in the range $50 \GeV \lesssim m_s \lesssim 500 \GeV$, reaching $750 \GeV$ close to its lower bound. For complex $S$, $m_s$ is more constrained, with values in the range 50-500 GeV. This can be easily understood since the dark matter density for a complex $S$ is twice that for a real $S$ of the same mass, therefore a smaller mass is required to produce the same density. From Fig.~\ref{msC} we see that a larger value of $\lambda_{s}(m_{t})$,
$\lambda_{s}(m_{t}) = 0.2$, permits a wider range of $S$ mass, with $m_{s}$ in the range 45~GeV to 1~TeV. If we instead considered the 2-$\sigma$ WMAP bound, the parameter space in Fig.~\ref{msA} and Fig.~\ref{msB} would increase, while Fig.~\ref{msC} would be unchanged.

 We note that while a large region of the allowed parameter space is at values of the $S$ mass which are large compared with the weak scale, there is no reason to expect the $S$ mass to be so large. The $S$ mass squared is $m_{s}^2 = \mso^2 + \lhs v^2/2$. Therefore if $\mso$ is of the order of the weak scale, which is the most natural possibility in a theory based on a single mass scale, we would expect $m_s$ to be no larger than a few hundred GeV.

There is a small region of the parameter space which satisfies $m_{s} < m_{h}/2$, with the lower bound on $m_{s}$ in the allowed region being slightly below $m_{h}/2$. This means that it is possible for the $S$ inflaton to be produced at the LHC via Higgs decay
\cite{bargerR}. Thermal relic $S$ dark matter would then originate from freeze-out of near resonant $S$ annihilation to $WW$ and $ZZ$ close to the Higgs pole. $m_s$ slightly below the Higgs pole also implies that $\lhs(m_{t})$ can be large, as can be seen from Fig. \ref{DMfig}. Therefore if $m_s < m_h/2$ then the $S$-nucleon scattering cross-section due to Higgs exchange is likely to be large, enhancing the possibility of observing $S$ dark matter in direct detection experiments.

Collider and direct dark matter detection experiments should be able to constrain the allowed parameter space\footnote{$\gamma$-ray and antimatter signals can also constrain the model \cite{yaguna}.}. Combined data from the D0 and CDF collaborations show that a Higgs boson mass in the range $160 \GeV < m_h < 170$~GeV is excluded at 95\% confidence level \cite{excl160170}.
This exclusion reduces the available parameter space of the model by a little less than a half, however it does not make a large difference to the range of $m_s$, as this is largest at low values of $m_h$. Present bounds on direct detection of $S$ dark matter from XENON10 and CDMSII rule out $S$ mass in the range 10 GeV to (50,70,75) GeV for Higgs masses (120,200,350) GeV \cite{ddbound,bargerR}. Comparing with Fig. 7, we see that the upper bound from direct detection is already close to the lower bound on the range of $m_s$ allowed by the $S$-inflation model. Thus although most of the parameter space is allowed at present, a substantial part of the ($m_s$, $m_{h}$) parameter space will be accessible to future dark matter detectors.

\section{Conclusions}
\label{conc}
We have shown that a gauge singlet scalar can serve simultaneously as the inflaton and as a thermal relic dark matter particle. As in the case of Higgs inflation, this requires a specific large non-minimal coupling of $S$ to gravity. Consistency of the model with (i) stability of the electroweak vacuum, (ii) perturbativity of the scalar potential as a function of $s$ and $h$ up to the Planck scale and (iii) the observed spectral index, constrains the $(\lhs(m_{t}), m_{h})$ parameter space. (The tensor-to-scalar ratio $r$ and the running of the spectral index $\alpha$ are both negligibly small compared with the observational limits.) Imposing the 5-year WMAP 1-$\sigma$ bound $n < 0.973$ implies that the range of the coupling of S to the Higgs is $|\lhs(m_{t})|\lesssim 0.15$, which can increase up to $|\lhs(m_{t})| \lesssim 0.55$ for small $S$ self-coupling and larger $n$. The range of Higgs masses is similar to but not identical to that of the Standard Model, with $145 \GeV \lesssim m_{h} \lesssim 170 \GeV$ for $n < 0.973$ and small $\lambda_{s}(m_{t})$, shifting to $130 \GeV \lesssim m_{h} \lesssim 170 \GeV$ for $n \gae 0.980$. Demanding that the $S$ annihilation rate through $\lhs(m_{t})$
produces the correct thermal relic $S$ dark matter density translates each $\lhs(m_{t})$ into a discrete set of possible values of $m_s$. Combined with the above constraints this determines a range of $m_s$ and $m_{h}$ which is simultaneously consistent with thermal relic dark matter and a stable, perturbative scalar potential which can account for the observed spectral index. The range of $m_s$ is sensitive to $\lambda_{s}(m_{t})$ and to whether $S$ is real or complex; for $\lambda_{s}(m_{t}) = 0.025$ and real $S$ the range is $50 \GeV \lesssim m_s \lesssim 750 \GeV$, with the upper limit increasing to 1 TeV for $\lambda_{s}(m_{t}) = 0.2$. For complex $S$ the range of $m_s$ is narrower, $50 \GeV \lesssim m_s  \lesssim 500 \GeV$ for  $\lambda_{s}(m_{t}) = 0.025$.

   Comparing with Higgs inflation in the unextended SM, a key difference is the range of possible $n$ versus $m_{h}$. In the Higgs inflation case this is expected to be very close to the classical value $n = 0.966$ except for $m_{h}$ close to the vacuum stability limit, $m_{h} \lesssim 130 \GeV$. In $S$ inflation a significant deviation from the classical value is expected over the whole range of $m_{h}$.  Planck is expected to be able to observe $n$ to an accuracy of $\pm 0.005$ (2-$\sigma$). Therefore Planck could provide evidence in favour of $S$ inflation relative to Higgs inflation, depending on what value of $m_{h}$ is observed at the LHC.

   Comparing with the model of \cite{clark}, which is based on the
same gauge singlet dark matter model but considers inflation along the Higgs direction, a notable difference is that in our model
$n$ is strictly larger than the classical value $n = 0.966$, whereas in \cite{clark} it is possible for the spectral index to become smaller than the classical value. In addition, the results of \cite{clark} indicate that the spectral index becomes close to or smaller than the classical value at $m_h \gae 160 \GeV$ (see Fig.~6 and 7 of \cite{clark}), whereas in our model the deviation from the classical value becomes larger as $m_h$ increases. Therefore it may be possible to distinguish between $S$-inflation and inflation along the Higgs direction, depending on the Higgs mass and the spectral index.

    There is a small region of the ($m_s, m_{h}$) parameter space with $m_s$ close to $m_{h}/2$ which is consistent with production of $S$ particles at the LHC via Higgs decay, $h \rightarrow SS$. Therefore if $S$ is observed at the LHC then thermal $S$ dark matter must originate from freeze-out of near resonant $S$ annihilations to $WW$ and $ZZ$ at the Higgs pole. This also allows $\lhs(m_{t})$ to be large and so a significantly large Higgs decay branching ratio may be expected. A large $S$-nucleon scattering cross-section via Higgs exchange is also expected in this case, which should allow the parameter space of the model to be probed by future dark matter detection experiments. The lower bound on $m_s$ is close to the present upper bound on $m_{s}$ from direct dark matter detectors (XENON10, CDMSII). Therefore significant constraints on the parameter space (or possibly direct detection of $S$ dark matter) may be expected in the future as dark matter detectors improve in sensitivity.

   It is natural to ask how predictive the $S$-inflation model can be. The barrier to a precisely predictive model is the dependence on the $S$ self-coupling $\ls(m_{t})$, which is not directly observable.
In principle, there are four observable quantities: $n$, $m_{h}$, $m_s$ and $\lhs(m_{t})$. The input parameters
of the $S$-inflation model are $m_{h}$, $m_s$, $\lhs(m_{t})$ and $\ls(m_{t})$. Therefore $n$ cannot be predicted exactly
as there will always be a dependence on $\ls(m_{t})$, even if the other parameters of the model are fixed by experiment.
Nevertheless, as we have shown, the possible range of $n$ can be constrained by vacuum stability and perturbativity constraints which constrain $\ls(m_{t})$. In addition, in the limit of small $\ls(m_{t})$ the model could become effectively independent of $\ls(m_{t})$. In this case we can predict $n$ if $m_{h}$, $m_s$ and
$\lhs(m_{t})$ are fixed by the LHC and direct dark matter experiments. If we are fortunate enough that $S$-inflation occurs in this limit, then the model can be predictive and testable.

    We have focused on the case of inflation along the $s$ direction. In general, inflation could occur along a more general
flat direction in the ($s$, $h$) plane, depending on the value of $\xi_{s}$, $\xi_{h}$ and the $V(s,h)$ scalar potential  couplings. $S$-inflation may be expected to be a good approximation in the limit $\xi_{s} \gg \xi_{h}$, assuming that the couplings $\lambda_{s}$, $\lambda_{h}$ and $\lhs$ are of a similar order of magnitude. The opposite limit of inflation along the $h$ direction was studied in \cite{clark}. This may be expected if $\xi_{h} \gg \xi_{s}$. The case of inflation along a general trajectory in the $s$, $h$ plane, which would be expected if $\xi_{s}$ and $\xi_{h}$ are the same order of magnitude, remains to be investigated.

     The $S$-inflation model provides a model for inflation and dark matter which is based purely on weak scale particles and interactions. In order to have a complete model of cosmology, we also need to address the issues of reheating and the origin of the baryon asymmetry. Reheating will be very similar to the case of Higgs inflation. In Higgs inflation reheating occurs via parametric resonance
of the oscillating Higgs field to $W$ bosons via the $|H|^2 |W|^2$ interaction \cite{shap2,preheating}.  In a similar way, in $S$-inflation reheating will occur via parametric resonance of $S$ oscillations to Higgs bosons via the $|S|^2 |H|^2$ interaction.
Baryogenesis could occur via the oscillating leptogenesis mechanism \cite{ol} or low-scale resonant leptogenesis \cite{rl} once the SM is extended by sterile neutrinos in order to account for neutrino masses.
Alternatively, baryogenesis could occur via electroweak baryogenesis, which may be possible in scalar extensions of the SM. Additional scalars interacting with the Higgs can produce a sufficiently strong 1st order electroweak phase transition. This usually requires that the gauge singlet scalar gains a vacuum expectation value (vev) after the transition \cite{barger2}, therefore a more complicated model with two or more additional scalars would be required \footnote{We note that it may be possible to evade this if the scalar has an expectation value prior to and during the electroweak phase transition but its vev vanishes in the vacuum after the transition \cite{mcdonald1}}.

In contrast to many inflation models, $S$-inflation is notable for the close relationship it implies between the observables of inflation (in particular, the spectral index), particle physics (in particular, the Higgs mass and Higgs decay width) and the direct detection of dark matter. It can therefore be directly tested by the experimental and observational advances which are anticipated in the near future as the LHC, Planck satellite and future direct dark matter detection experiments come to fruition.

\section*{Acknowledgements}

This work was supported by the European Union through the Marie Curie Research and Training Network "UniverseNet" (MRTN-CT-2006-035863). The work of R.L. was also supported by STFC.

\section*{Appendix A: Renormalization Group Equations}

 \renewcommand{\theequation}{A-\arabic{equation}}
 \setcounter{equation}{0}

In this appendix we relate the gauge singlet model to the notation of \cite{mv} used to compute the RG equations. We also briefly review the computation of the RG equations for $\xi_{s}$ and $\xi_{h}$.

\subsection{RG equations for scalar couplings}

In \cite{mv} the general RG equations are given to two-loops in the $\overline{{\rm MS}}$ scheme. The anomalous dimensions and $\beta$-functions are expressed in terms of real (reducible) representations of the scalar fields and Majorana spinors. To obtain the modification of the RG equations due to the $S$ field, we express the Higgs doublet and gauge singlet scalars as a set of six real scalar fields, $\phi_{i} \; (i = 1...6)$, where
\be{ba1} H = \frac{1}{\sqrt{2}} \left( \begin{array}{c} \phi_{1} + i \phi_{2} \\ \phi_{3} + i \phi_{4} \end{array} \right) \ee
and
\be{ba2} S =  \frac{1}{\sqrt{2}} \left( \phi_{5} + i \phi_{6} \right)  ~.\ee
(For the case of real $S$, $\phi_{6} = 0$.)

  Writing the Higgs doublet as a real representation in the form $(\phi_{1},\phi_{2},\phi_{3},\phi_{4})^{T}$, the
$SU(2)_{L}$ generators ($\theta^{A}_{ab}$ in the notation of \cite{mv}) are
\be{b1} \theta^{1} = \frac{1}{2} \left(
\begin{array}{cccc}
0 & 0 & 0 & i \\
0 & 0 & -i & 0 \\
0 & i & 0 & 0 \\
-i & 0 & 0 & 0
\end{array}
\right)   ~,\ee
\be{b2} \theta^{2} = \frac{1}{2} \left(
\begin{array}{cccc}
0 & 0 & -i & 0 \\
0 & 0 & 0 & -i \\
i & 0 & 0 & 0 \\
0 & i & 0 & 0
\end{array}
\right)   ~,\ee
and
\be{b3} \theta^{3} = \frac{1}{2} \left(
\begin{array}{cccc}
0 & i & 0 & 0 \\
-i & 0 & 0 & 0 \\
0 & 0 & 0 & -i \\
0 & 0 & i & 0
\end{array}
\right)   ~.\ee
The $U(1)_{Y}$ generator is
\be{b4} \theta^{Y} = i \left(
\begin{array}{cccc}
0 & Y & 0 & 0 \\
-Y & 0 & 0 & 0 \\
0 & 0 & 0 & Y \\
0 & 0 & -Y & 0
\end{array}
\right)   \ee
where $Y = 1/2$ is the hypercharge of the complex fields in the Higgs doublet.

   The only Yukawa coupling we consider is the top quark Yukawa coupling. In 4-component spinor notation this is (in the notation of \cite{mv})
\be{b5} \overline{q} \underline{H} \phi^{\dagger\;c} q  + h.c.    ~,\ee
where $\underline{H}$ is the Yukawa coupling matrix, $q = (u_{L}, d_{L})^{T}$ is the $SU(2)_{L}$ quark doublet
and $\phi$ is the Higgs doublet. In our case
\be{b6} \overline{q} \underline{H} \phi^{\dagger\;c} q  \equiv \overline{t}_{R} y_{t} t_L \phi^{0} - \overline{t}_{R} y_{t} b_L \phi^{+}   ~.\ee
(In this we have suppressed colour indices.)
We define a reducible representation $\psi_{i}$ (in the notation of \cite{mv}) by $(\psi_{1},\psi_{2},\psi_{3}) = (t_{R}^{c}, t_{L}, b_{L})$, where $t_{L},\; b_{L}$ and $t_{R}^{c}$ are the two-component spinors which form the Dirac spinors in the chiral representation $\left(t \equiv (t_{L}, t_{R})^{T}~\mbox{etc}\right)$, with $t_{R}^{c} = -i \sigma_{2} t_{R}^{*}$. The
Yukawa coupling can then be written as
\be{b7} Y^{a}_{ij} \psi_{i} \xi \psi_{j} \phi_{a} + h.c.\;\;\;\; \left(a = 1,2,3,4\right)\ee
where
\be{b8} Y^{1} = \frac{1}{\sqrt{2}}
\left(
\begin{array}{ccc}
0 & y_{t} & 0 \\
y_{t} & 0 & 0 \\
0 & 0 & 0
\end{array}
\right)   ~,\ee
\be{b9} Y^{2} = \frac{i}{\sqrt{2}}
\left(
\begin{array}{ccc}
0 & y_{t} & 0 \\
y_{t} & 0 & 0 \\
0 & 0 & 0
\end{array}
\right)   ~,\ee
\be{b10} Y^{3} = \frac{1}{\sqrt{2}}
\left(
\begin{array}{ccc}
0 & 0 & -y_{t} \\
0 & 0 & 0 \\
-y_{t} & 0 & 0
\end{array}
\right)   \ee
and
\be{b11} Y^{4} = \frac{i}{\sqrt{2}}
\left(
\begin{array}{ccc}
0 & 0 & -y_{t} \\
0 & 0 & 0 \\
-y_{t} & 0 & 0
\end{array}
\right)   ~.\ee
The corresponding $SU(2)_{L}$ generators $t^{A}$ acting on
$\psi$ are
\be{b12} t^{1} = \frac{1}{2}
\left(
\begin{array}{ccc}
0 & 0 & 0 \\
0 & 0 & 1 \\
0 & 1 & 0
\end{array}
\right)   ~,\ee
\be{b13} t^{2} = \frac{1}{2}
\left(
\begin{array}{ccc}
0 & 0 & 0 \\
0 & 0 & -i \\
0 & i & 0
\end{array}
\right)   ~\ee
and
\be{b14} t^{3} = \frac{1}{2}
\left(
\begin{array}{ccc}
0 & 0 & 0 \\
0 & 1 & 0 \\
0 & 0 & -1
\end{array}
\right)  ~. \ee
The $U(1)_{Y}$ generator is
\be{b15} t^{Y} =
\left(
\begin{array}{ccc}
-\frac{2}{3} & 0 & 0 \\
0 & \frac{1}{6}  & 0 \\
0 & 0 &  \frac{1}{6}
\end{array}
\right)   ~.\ee
(Suppressed colour indices should be summed over when taking traces in the formulae of \cite{mv}.)
Finally, $\kappa = 1/2$ should be used in \cite{mv} since $\psi_{i}$ are two-component spinors.
With these definitions of $\theta^{A}$, $Y^{a}$ and $t^{A}$, the formulae in \cite{mv} can be used to compute the RG equations
to two-loop order as a function of the t-quark Yukawa coupling, gauge couplings and the scalar couplings.

\subsection{RG equations for $\xi_s$ and $\xi_h$}

The 1-loop RG equations for the non-minimal gravity couplings $\xi_{h}$ and $\xi_{s}$ are obtained as follows. For a general theory of scalars $\phi_{i}$ with mass terms and non-minimal couplings in the Lagrangian
\be{b16} {\cal L}  \supset \frac{1}{2}m_{ij}\phi_{i} \phi_{j}  +  \frac{1}{2}\xi_{ij}\phi_{i} \phi_{j}    ~,\ee
the 1-loop bare and renormalized $\xi_{ij}$ are related by \cite{od}
\be{b17} \xi_{o\;ij}  = \left(\xi_{kl} - \frac{1}{6} \delta_{kl}\right)Z_{2\;ij}^{kl} + \frac{1}{6} \delta_{ij}  ~,\ee
where $Z_{2\;ij}^{kl}$ is the mass renormalization,
\be{b18} m_{o\;ij}^{2} = Z_{2\;ij}^{kl} m_{kl}^2   ~.\ee
Therefore the RG equations for $\xi_{ij}$ are related to the mass anomalous dimensions $\gamma_{m\;ij}^{kl}$ by
\be{b19}  \mu \frac{d \xi_{ij}}{d \mu} = \left( \xi_{mn} - \frac{1}{6} \delta_{mn}\right) \gamma_{m\;ij}^{kl}   ~.\ee
$\gamma_{\;m\;\;ij}^{\;kl}$ can be easily derived by applying the scalar potential RG equations to the 1-loop effective potential in order to obtain the $\beta$-function of the mass term \cite{Ford:1992mv}, $\beta_{m^2\;ij} \equiv \gamma_{m\;ij}^{ab} m^{2}_{ab}$.

Finally, the RG equations must be modified at large $s$ or $h$ by suppressing the propagator for the corresponding real scalar field. Note that one does not suppress all the components of the Higgs doublet for the $h$ direction nor the complex component of $S$ in the $s$ direction.

\section*{Appendix B: Gauge Singlet Scalar Dark Matter Density}

\renewcommand{\theequation}{B-\arabic{equation}}
 \setcounter{equation}{0}

  In this appendix we give the $S$ annihilation cross-section times relative velocity, $\langle\sigma v_{rel}\rangle$, and the resulting dark matter density. We will approximate $\langle\sigma v_{rel}\rangle$ by the centre-of-mass cross-section for non-relativistic $S$ annihilation. The cross-sections for real and complex $S$ are the same; we will present results for the real case. The tree-level processes contributing to $S$ annihilation are (i) $SS \rightarrow hh$, (ii) $SS \rightarrow WW$, (iii) $SS \rightarrow ZZ$ and (iv) $SS \rightarrow \overline{f}f$ (where $f$ is a Standard Model fermion). (i) proceeds via a 4-point contact interaction, an s-channel Higgs exchange interaction and a t- and u-channel $S$ exchange interaction. The resulting $\langle\sigma v_{rel}\rangle$ is
\bea \langle\sigma v_{rel}\rangle_{hh}& =& \frac{\lhs^{2}}{64 \pi m_s^{2}}
\left[ 1 + \frac{3 m_{h}^2}{\left(4 m_s^2 - m_h^2\right) } + \frac{2 \lhs v^2}{\left(m_h^2 - 2 m_s^2\right) }\right]^2 \nonumber \\
& & \times \left(1-\frac{m_{h}^{2}}{m_s^{2}}\right)^{1/2}~.
\eea
$SS \rightarrow WW,\;ZZ,\;\overline{f}f$ all proceed via s-channel Higgs exchange. The corresponding
$\langle\sigma v_{rel}\rangle$ are:
\bea
\langle\sigma v_{rel}\rangle_{WW}& =& 2\Bigg(1+\frac{1}{2}\left(1-\frac{2 m_s^{2}}{m_{W}^{2}}\right)^{2} \Bigg) \left(1-\frac{m_{W}^{2}}{m_s^{2}}\right)^{1/2} \nonumber \\
& & \times \frac{\lhs^{2} m_{W}^{4}}{8 \pi m_s^{2}\left(\left(4 m_s^{2} -m_{h}^{2}\right)^{2}+m_{h}^{2}\Gamma_{h}^{2}\right)  }
~,\eea
\bea
\langle\sigma v_{rel}\rangle_{ZZ}& =&  2\Bigg(1+\frac{1}{2} \left(1-\frac{2 m_s^{2}}{m_{Z}^{2}} \right)^{2}\Bigg) \left(1-\frac{m_{Z}^{2}}{m_s^{2}}\right)^{1/2} \nonumber \\
& & \times \frac{\lhs^{2} m_{Z}^{4}}{16 \pi m_s^{2} \left(\left(4 m_s^{2} -m_{h}^{2}\right)^{2}+m_{h}^{2}\Gamma_{h}^{2}\right)  }
\eea
and
\be{a4}  \langle \sigma v_{rel}\rangle_{\overline{f}f} =\frac{m_{W}^{2}}{\pi g^{2}} \frac{\lambda_{f}^{2}
\lhs^{2} }{\left(\left(4m_s^{2}-m_{h}^{2}\right)^{2}+m_{h}^{2}\Gamma_{h}^{2}\right) } \Bigg(1-\frac{m_{f}^{2}}{m_s^{2}}\Bigg)^{3/2}
~.\ee
Here the fermion Yukawa coupling is $\rm \lambda_{f} =
m_{f}/v$ where $v = 246.22$~GeV and $\rm m_{f}$ is the fermion
mass. $\rm
\Gamma_{h}$ is the Higgs decay width. (Fermions should be summed over colours.)

   The dark matter density is calculated using the Lee-Weinberg approximation.
For real $S$ the present total mass
density in S scalars is \cite{mcdonald2}
\be{a5} \Omega_{S} \equiv
\frac{\rho_{S}}{\rho_{c}} =
\frac{g(T_{\gamma})}{g(T_{fS})} \frac{K}{T_{\gamma}
x_{fS}
\langle\sigma_{ann} v_{rel}\rangle
}\left(\frac{T_{\gamma}^{4}}{\rho_{c}}\right)
\frac{\left(1-3x_{fS}/2\right)}{\left(1-x_{fS}/2\right)}
~,\ee
where $T_{fS}$ is the $S$ freeze-out temperature, $x_{fS} = T_{fS}/m_s$ and
$K = \left(4 \pi^{3} g(T_{fS})/45 M_{Pl}^{2} \right)^{1/2}$ where $g(T)$ is the effective number
of relativistic degrees of freedom. The density for complex $S$ is twice that for real $S$,
due to the additional degree of freedom.

\end{document}